\documentclass[twocolumn,twocolappendix]{aastex631}

\usepackage{amsmath,amssymb}
\usepackage{physics}
\usepackage{amsfonts}
\usepackage{mathtools}

\usepackage{etoolbox}
\usepackage[refpage]{nomencl}

\providetoggle{nomsort}
\settoggle{nomsort}{true} 

\makeatletter
\iftoggle{nomsort}{%
    \let\old@@@nomenclature=\@@@nomenclature        
        \newcounter{@nomcount} \setcounter{@nomcount}{0}%
        \newcommand{\threedigits}[1]{\ifnum#1<100 0\two@digits{#1} \else \number#1\fi}
        \renewcommand\the@nomcount{\threedigits{\value{@nomcount}}}
        \def\@@@nomenclature[#1]#2#3{
          \addtocounter{@nomcount}{1}%
        \def\@tempa{#2}\def\@tempb{#3}%
          \protected@write\@nomenclaturefile{}%
          {\string\nomenclatureentry{\the@nomcount\nom@verb\@tempa @[{\nom@verb\@tempa}]%
          \begingroup\nom@verb\@tempb\protect\nomeqref{\theequation}%
          |nompageref}{\thepage}}%
          \endgroup
          \@esphack}%
      }{}
\makeatother

\newcommand{\mynomone}[3][section]{%
  \begingroup\edef\x{\endgroup
  \unexpanded{\nomenclature{#2}}%
    {\unexpanded{#3} \hspace*{\fill}  (\csname the#1\endcsname)}}\x}

\newcommand{\mynomtwo}[4][section]{%
  \begingroup\edef\x{\endgroup
  \unexpanded{\nomenclature[#2]{#3}}%
    {\unexpanded{#4} \hspace*{\fill}  (\csname the#1\endcsname)}}\x}

\renewcommand\nomgroup[1]{%
  \item[\bfseries
  \ifstrequal{#1}{A}{Acronyms}{%
  \ifstrequal{#1}{S}{Symbols}{%
  \ifstrequal{#1}{C}{Other Symbols}{}}}%
]}

\newcounter{logglabel}

\newcommand{\mynom}[3][S]{\nomenclature[#1]{#2}{#3~}}

\makenomenclature

\newcommand{\eq}[1]{Eq.~(\ref{#1})}

\shorttitle{Tides on Lava Worlds}
\shortauthors{Farhat et al.}
\graphicspath{{./}{figures/}}

\begin{document}

\title{Tides on Lava Worlds:\\ Application to Close-in Exoplanets and the Early Earth-Moon System}

\author[0000-0001-7864-6627]{Mohammad Farhat}
\altaffiliation{Miller Fellow}
\affiliation{IMCCE, CNRS, Observatoire de Paris, PSL University, Sorbonne Université, 77 Avenue Denfert-Rochereau, 75014, Paris, France }
\affiliation{Department of Astronomy, University of California, Berkeley, Berkeley, CA 94720-3411, USA  }
\affiliation{Department of Earth and Planetary Science, University of California, Berkeley, Berkeley, CA 94720-4767, USA}
\author[0000-0002-9577-2489]{Pierre Auclair-Desrotour}
\affiliation{IMCCE, CNRS, Observatoire de Paris, PSL University, Sorbonne Université, 77 Avenue Denfert-Rochereau, 75014, Paris, France }

\author[0000-0002-5057-7743]{Gwena{\"e}l Bou{\'e}}
\affiliation{IMCCE, CNRS, Observatoire de Paris, PSL University, Sorbonne Université, 77 Avenue Denfert-Rochereau, 75014, Paris, France }

\author[0000-0002-3286-7683]{Tim Lichtenberg}
\affiliation{Kapteyn Astronomical Institute, University of Groningen, P.O. Box 800, 9700 AV Groningen, The Netherlands}

\author[0000-0003-2634-789X]{Jacques Laskar}
\affiliation{IMCCE, CNRS, Observatoire de Paris, PSL University, Sorbonne Université, 77 Avenue Denfert-Rochereau, 75014, Paris, France }



\begin{abstract}
Understanding the physics of planetary magma oceans has been the subject of growing efforts, in light of the increasing abundance of Solar system samples and extrasolar surveys. A rocky planet harboring such an ocean is likely to interact tidally with its host star, planetary companions, or satellites. To date, however, models of the tidal response and heat generation of magma oceans have been restricted to the framework of weakly viscous solids, ignoring the dynamical fluid behavior of the ocean beyond a critical melt fraction. Here we provide a handy analytical model that accommodates this phase transition, allowing for a physical estimation of the tidal response of lava worlds. We apply the model in two settings: The tidal history of the early Earth-Moon system in the aftermath of the giant impact; and the tidal interplay between short-period exoplanets and their host stars. For the former, we show that the fluid behavior of the Earth's molten surface drives efficient early Lunar recession to ${\sim} 25$ Earth radii within $10^4{-} 10^5$ years, in contrast with earlier predictions. For close-in exoplanets, we report on how their molten surfaces significantly change their spin-orbit dynamics, allowing them to evade spin-orbit resonances and accelerating their track towards tidal synchronization from a Gyr to Myr timescale. Moreover, we re-evaluate the energy budgets of detected close-in exoplanets, highlighting how  the surface thermodynamics of these planets are likely controlled by enhanced, fluid-driven tidal heating, rather than vigorous insolation, and how this regime change substantially alters predictions for their surface temperatures.

\end{abstract}



\section{Introduction} \label{sec:intro}

Magma oceans are thought to be a ubiquitous feature of the growing planetary diversity. Physical pathways of their formation span several sources of heat generation, including: the extremely energetic accretionary impacts late in the planetary growth stage \citep[e.g.,][]{abe1985formation,stevenson1987origin,benz1990terrestrial,tonks1993magma,elkins2012magma,nakajima2015melting}{}{}; strong stellar irradiation impinging upon the surfaces of close-in planets, driving their temperatures beyond the melting limit of most mineral species \citep[e.g.,][]{leger2009transiting,leger2011extreme,demory2016map}{}{}; shear heating during core formation, as gravitational potential energy is dissipated while iron-rich materials are drained to the center of the differentiating planet \citep[e.g.,][]{solomon1979formation,ricard2009multi,rubie2007formation}{}{}; induction heating around strongly magnetized stars \citep[e.g.,][]{kislyakova2017magma,kislyakova2018effective}; and possibly, intense interior tidal heating for planets residing on sustained eccentric orbits or trapped in mean motion resonances \citep[akin to Jupiter's Io,][]{peale1979melting,2018arXiv180405110H}{}{}.

Understanding the underlying physics of magma oceans is essential to address a multitude of questions pertaining to rocky planetary formation and evolution. In the inner Solar system, early magma oceans are modeled to determine, for instance, the redox states of the formed mantles and their overlying atmospheres \citep[e.g.,][]{deng2020magma,sossi2020redox,hirschmann2022}{}{}, interior differentiation and core formation \citep[e.g.,][]{ichikawa2010direct,hirose2017crystallization,davies2020transfer}{}{}, scenarios of crustal formation and stability \citep[e.g.,][]{bouvier2018evidence,michaut2022formation}{}{}, and mechanisms of plate tectonic initiation and evolution \citep[e.g.,][]{foley2014initiation,schaefer2018magma}{}{}. Different pathways of such mechanisms have certainly contributed to the diversity of the inner Solar system, eventually prescribing  the (non-)favorable habitability conditions.  

Beyond our Solar system, close-in exoplanets harboring (or expected to harbor) magma oceans are prioritized targets for characterization due to their favorable orbital proximity and strong thermal emission \citep{2018PASP..130k4401K,lichtenberg2024}. In such settings, the star-planet strong interaction is expected to deplete the planet's atmospheric reservoir of volatiles over time \citep{2019AREPS..47...67O}. In response, the magma ocean is expected to vaporize and outgas, rendering the atmosphere heavy and tenuous with a partly vaporized rock composition \citep[e.g.,][]{schaefer2009chemistry,miguel2011compositions,perez2013catastrophic,zieba2022k2}{}{}. Recently, for instance, JWST observations of the super-Earth lava world 55~Cancri~e indicated first potential signs of 
CO/CO$_2$ in the atmosphere \citep[][]{hu2024secondary}{}{}. At equilibrium then, characterizing the atmospheres overlying magma oceans can provide a novel window into understanding exoplanetary interior compositions and properties \citep[e.g.,][]{ito2015theoretical,kite2016atmosphere,dorn2021hidden,wordsworth2022atmospheres}{}{}. 
 
Since multiple magma ocean formation pathways are associated with the planet's interaction with companions, tidal interactions in the system are of significance. As the planet occupies its orbit, a differential gravitational forcing of any companion (be it the host star, another planet, or a satellite), will drive flexure in the planet, inducing a tidal bulge that can oscillate in amplitude. If the planet is of solid composition, the forced  tidal flexure of the interior is resisted by friction in the mantle, releasing heat in the process \citep[see e.g.,][for a recent review]{bagheri2022tidal}{}{}. For planets hosting fluid oceans and atmospheres, the tidal response associated with these layers is driven by different dynamical processes, and is likely to be more significant than that of the solid counterpart, as energy dissipation pathways are more efficient and islands of dynamical resonances are accessible \citep[e.g.,][]{auclair2017atmospheric,auclair2018oceanic}{}{}. 

Interestingly, a magma ocean, be it global or sporadic, transient or persistent, residing on the surface or underneath, lives at and around the rheological transition between solid and fluid phases. As such, with the layer evolving across the phases, its tidal response should follow the transition, the latter being commonly parameterized by the fraction of melt in the ocean. To this end, however, modeling the tidal response of a magma ocean has been restricted in earlier studies to the framework of solid rheologies, whereby the phase transition is accounted for by an adjustment of physical parameters (typically the viscosity and shear modulus), instead of changing the dynamical medium and its governing equations of the tidal flow \citep[e.g.,][]{zahnle2015tethered,chen2016tidal,barth2021magma,korenaga2023rapid,downey2023thermal}{}{}.   

In the solid response framework, the amount of tidal heating generated in the magma ocean is controlled by the mantle's temperature-dependent viscosity: as the temperature increases, more melt is generated, the viscosity decreases, and so does the generated heat. It is thus interesting to investigate how the fluid behavior of the magma ocean, which is triggered as temperature increases, can alter this classical picture. To our knowledge, the work of \citet[][]{tyler2015tidal}{}{} offers the only study which focuses on the fluid tidal behavior in magma oceans. The motivation of \citet[][]{tyler2015tidal}{}{} was to explain the  observable signatures of the putative subsurface magma ocean on Jupiter's Io: its enhanced dissipation and surface tidal heating patterns. Here, we aim to expand on the theoretical modeling end in this direction by developing an analytical model that captures the tidal response of a rocky planet harboring a magma ocean. 

\begin{figure}[t]
\includegraphics[width=.47\textwidth]{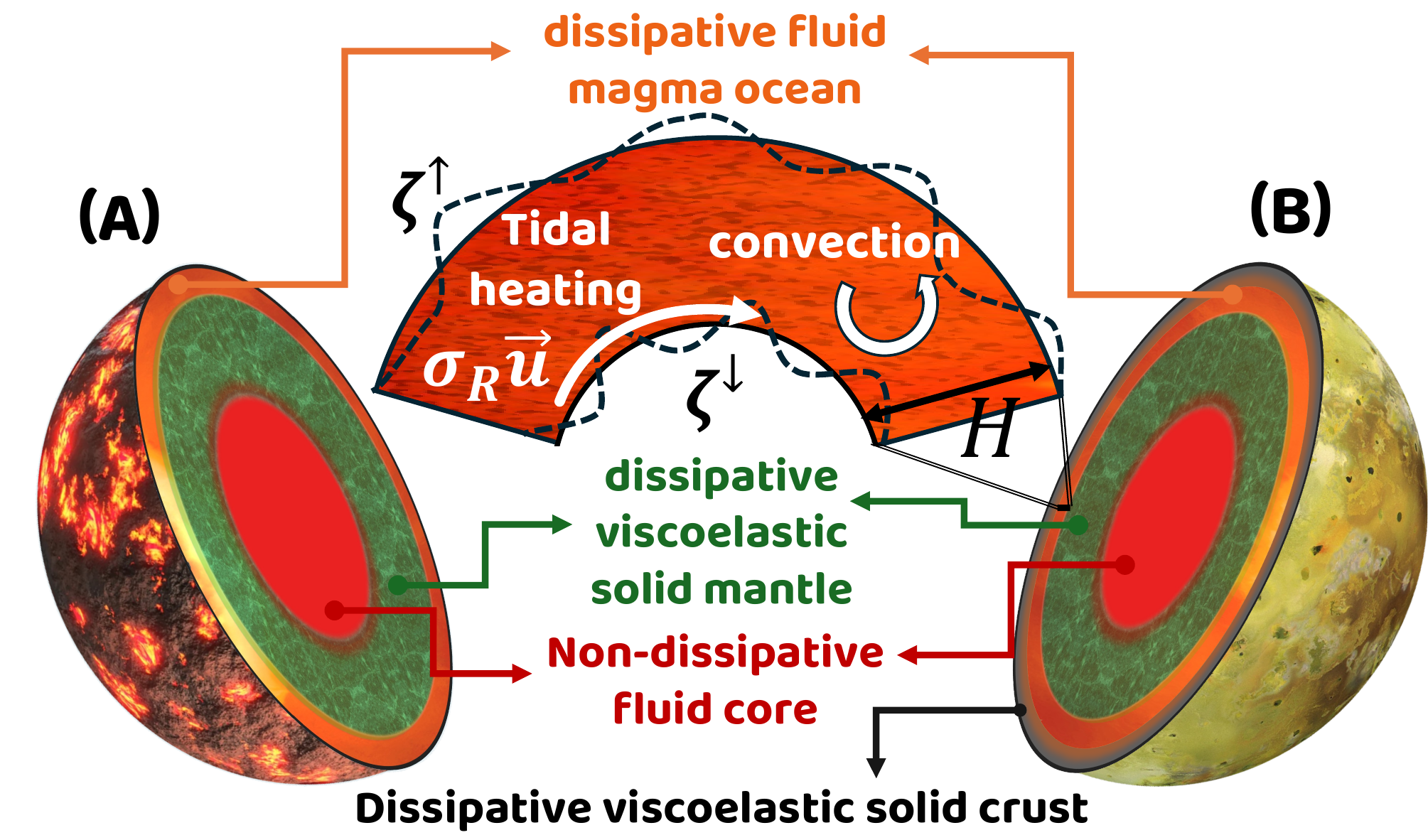}
\caption{{A schematic illustrating the different planetary models under study. In both models, (A) and (B), the planet has a molten core that contributes negligibly to the tidally dissipated energy (this also stands if the core is differentiated into a solid inner and fluid outer parts). On top of the core lies a viscoelastic solid mantle which can be stratified and further divided into layers of uniform physical properties. Tidal dissipation in this mantle is computed in the solid tidal formalism using \texttt{ALMA3} (Section \ref{Section_Tides_Solid}). The solid mantle is enveloped by a fluid magma ocean of unperturbed uniform thickness $H$, which is the focus of our study. Under tidal forcing of the ocean and the mantle, the ocean's bottom (top) is displaced vertically by $\zeta^{\downarrow}$ ($\zeta^{\uparrow})$. This forcing also leads to heating (by virtue of several dissipative mechanisms in reality, but mainly due to friction at the ocean-solid mantle interface, as shown in the sketch). The heating is counteracted by convective cooling.  This molten layer can either be the outermost layer of the planet (A), or it can live under a solidified crust (B). } }
\label{magma_sketch}
\end{figure}

{Surprisingly, there is yet to be a unified, cross-disciplinary definition of magma, magma oceans, and lava worlds \citep[see e.g.,][]{SOLOMATOV201581,chao2021lava,christiansen2022origin}. In our model hereafter, we define a magma ocean as a layer where the melt fraction is significant enough such that the layer, as a whole, behaves rheologically as a fluid, in the presence of suspended solids \citep[e.g.,][]{tonks1993magma,abe1997thermal}. This is in contrast to the case of a magma sponge layer with low melt fraction where the fluid percolates through a matrix of solid and crystal mush, rendering the behavior of the layer solid-like. The critical melt fraction at which this rheological transition occurs ranges, as suggested by experiments, between 0.2 and 0.6 for olivine and silicates \citep[e.g.,][]{abe1993thermal,scott2006effect,costa2009model}. We shall study the response of such a layer living on the surface of a planet or underneath a solid crust. In both cases, we shall also self-consistently couple the response of this layer to the response of a solid mantle underneath. We provide a sketch of our modeled planet in Figure \ref{magma_sketch}. } 

In Section \ref{Sec:Self_consistent_tidal_response} we describe the used methods and the analytical framework of our model, arriving at a closed-form solution for the planetary Love number, which captures the tidal response of a planet harboring a magma ocean. In Section \ref{Section_Early_Earth_Moon}, we  adopt the developed model to investigate the early tidal evolution of the Earth-Moon system, with the Earth partially molten after the Moon forming giant impact. In Section \ref{Section_Close_In_Exo}, we explore the implications of fluid-driven tidal heating on exoplanets: their spin-orbit dynamics and surface energy budgets, leading to re-analyzing the predicted surface temperatures of detected close-in exoplanets. We then summarize, highlighting the limitations of our developed model and discussing possible directions for future work in Section \ref{Section_discussion}.

\section{The Self-Consistent Planetary Tidal Response}\label{Sec:Self_consistent_tidal_response}
\subsection{The Tidal Response of the Solid Interior}\label{Section_Tides_Solid}

We first focus on the solid interior of the planet {(the viscoelastic solid mantle part in Figure \ref{magma_sketch})}, aiming to recover the tidal response of a spherically symmetric,  radially multi-layered, self-gravitating viscoelastic mantle by computing the viscoelastic Love numbers. The latter are traditionally obtained within the framework of the viscoelastic normal modes formalism, first introduced by \cite{peltier1974impulse}, in which a coupled system of differential equations describing mantle deformation under loading or tidal stresses is solved \citep[see e.g.,][]{sabadini2016global}. In practice, solving the said system has been achieved by direct numerical integration \citep[e.g.,][]{tobie2005tidal}, or by using standard propagation methods across the spherical shells \citep[e.g.,][]{wu1982viscous,vermeersen1996compressible}. The deformation solution, and consequently the Love numbers, are first obtained in the frequency domain, by virtue of the correspondence principle\footnote{The principle entails that if a deformation solution in the linear elastic limit is known, the solution of the analogous problem of a linear viscoelastic medium is readily obtained by replacing all time-dependent quantities by their frequency-dependent transforms.} \citep{biot1956theory}, then transformed to the time domain via classical Laplace or Fourier inverse transformations.  However, certain limitations to the traditional inversion approaches emerge when the stratification of the mantle or the complexity of the rheological model are increased \citep[][]{spada2006using,farhat2022constraining}.

Here, we use the code \texttt{ALMA3} \citep{spada2006using,spada2008alma,melini2022computing}, which performs the Laplace inversion numerically using the Post–Widder formula to obtain the deformation solution and the Love numbers \citep{post1930generalized,widder1934inversion}. Using \texttt{ALMA3}, we compute, for a given harmonic degree, $n$, the tidal Love numbers $\{h_n^{T},l_n^{T},k_n^{T}\}$\mynom{$h_n^{T},l_n^{T},k_n^{T}$}{tidal Love numbers} and the loading Love numbers $\{h_n^{L},l_n^{L},k_n^{L}\}$\mynom{$h_n^{L},l_n^{L},k_n^{L}$}{loading Love numbers}. \texttt{ALMA3} takes as an input a planetary interior model with a prescribed number of layers enveloping a central core. We characterize each layer by a rheological law, uniform density, $\rho$, shear modulus, $\mu$, and viscosity, $\eta$, while we model the core to respond to stresses as an inviscid fluid. As a default setup, and unless stated otherwise, we adopt an Andrade rheology for the solid parts of the planet when computing the Love number using \texttt{ALMA3}. The study of transient creep of most materials at high temperatures reveals that they abide by the Andrade rheological law \citep[e.g.,][]{louchet2009andrade}{}{}. Moreover, the analysis of geophysical data on Earth pertaining to seismic waves, the Chandler wobble, and mantle normal modes also support an Andrade rheological description \citep[e.g.,][]{anderson1979frequency,Castillo-Rogez,efroimsky2012bodily,gevorgyan2020andrade}{}{}.

{For an Andrade rheology, the time-dependent creep function, $J(t)$, relating the material strain to the tidal stress, reads as:
\begin{equation}
    J(t) = \frac{1}{\mu}\left[1+\left(\frac{t}{\tau_{\rm A}}\right)^\alpha + \frac{t}{\tau_{\rm M}}\right].
\end{equation} 
With this definition, the solid Andrade flow describes an elastic regime at high forcing frequencies (for $t\rightarrow0)$, characterized by the inverse of unrelaxed rigidity, $1/\mu$; a viscous regime at low frequencies, characterized by the viscoelastic Maxwell timescale $\tau_{\rm M}=\eta/\mu$; and a transient anelastic transition in between, characterized by the Andrade timescale, $\tau_{\rm A}$, and parameterized by the exponent of the creep strain law, $\alpha$. As such, in contrast to the often-adopted Maxwell rheology, which lacks the transient element, the Andrade rheology attenuates the rapid decay of the anelastic component of the deformable solid for high viscosities, or equivalently, at high tidal forcing frequencies, which is a relevant regime for this study as will be shown later.  The dimensionless exponent, $\alpha$ is an empirical constant that takes values between 0.2 and 0.4 for silicates and ices, and between 0.14 and 0.2 for partial melts \citep[e.g.,][]{efroimsky2012bodily}. For the Andrade timescale, $\tau_{\rm A}$, it is found that  $\tau_{\rm A}\geq \tau_{\rm M}$ \citep[][]{Castillo-Rogez,makarov2012conditions}{}{}, with values on the shorter end pronouncing the effect of anelasticity, and larger values weakening it.}

\subsection{The Tidal Response of an isolated Magma Ocean in the Fluid Limit}\label{Section_Magma_isolated}
On a planet with radius $R_{\rm p}$, we are interested in computing the tidal response of a magma layer occupying a spherical shell of uniform thickness, $H$. The rheology of this layer, whether it resides on the surface of the planet or underneath, is strongly dependent on its melt fraction, $F_{\rm m}$. A rheological transition occurs when $F_{\rm m}$ exceeds a critical melt fraction, $F_{\rm m, c}$, rendering the behavior of the magma layer under tidal stresses a fluid-like behavior, rather than a solid one {\citep[traditionally, $F_{\rm m, c}$ is set to 0.4, with experiments suggesting a range from 0.2 to 0.6 for olivine and silicates;][]{abe1993thermal,scott2006effect,costa2009model}}. This transition is conventionally accounted for by an adjustment of the layer parameters, namely the viscosity, which drops several orders of magnitudes from a solid layer viscosity to a melt viscosity \citep[e.g.,][]{solomatov2007magma,zahnle2015tethered}{}{}. Our main aim here is to investigate how this modeling choice leaves out significant tidal features that can only be captured when fluid dynamics are properly implemented. 

For a magma layer having $F_{\rm m}\geq F_{\rm m, c}$, we employ the Laplace Tidal Equations (LTEs), which describe tidal dynamics in relatively thin shell fluid layers \citep[e.g.,][]{longuet1968eigenfunctions}. The LTEs have been used semi-analytically to study tidal dynamics within global surface oceans \citep[e.g.,][]{auclair2018oceanic,auclair2019final,motoyama2020tidal,tyler2021tidal}, hemispherical or spherical cap surface oceans \citep[e.g.,][]{webb1980tides,farhat2022resonant,auclair2023can}, subsurface oceans \citep[e.g.,][]{tyler2011tidal,matsuyama2014tidal}, and atmospheres \citep[e.g.,][]{siebert1961atmospheric,lindzen1972lamb,auclair2017atmospheric,farhat2024thermal}. In the framework of the LTEs, with nonlinearities neglected, and with the mean flows averaged out, momentum and mass conservation read as:
\begin{subequations}
\label{momentum_continuity}
\begin{align}\label{momentum1}
&\partial_t\Vec{u}+\sigma_{\rm R}\Vec{u}+\Vec{f}\cross\Vec{u}+g\grad\zeta = g\grad\zeta_\mathrm{eq},\\
&\partial_t\zeta +\grad\cdot \left( H\Vec{u} \right)=0. \label{continuity1}
\end{align}
\end{subequations}
Here, $t$\mynom{$t$}{time coordinate} is the time coordinate and $\partial_t=\partial/\partial t$, $\Vec{u}= u_\theta\hat{\theta}+u_\lambda \hat{\lambda}$ is the horizontal velocity field of the tidal flow in polar coordinates [$\theta$ being the co-latitude and $\lambda$ the longitude, and $(\hat{\theta}$,$\hat{\lambda})$ being their respective unit vectors];  $g$ is the gravitational acceleration at the surface, $\zeta$ is the tidally varying fluid depth, $\zeta_\mathrm{eq}$ is the equilibrium depth variation, $\grad=R_{\rm p}^{-1}\left[\hat{\theta}\partial_{\theta}+\hat{\lambda}(\sin\theta)^{-1}\partial_{\lambda}\right]$ is the two-dimensional horizontal gradient operator, and $\grad\cdot\vec u = (R_{\rm p}\sin\theta)^{-1}\left[\partial_\theta(\sin\theta  u_\theta)  + \partial_\lambda u_\lambda \right]$ is the horizontal velocity divergence; $f=2\Omega\cos\theta$ is the Coriolis parameter for a planet rotating with velocity $\Omega$, and $\sigma_{\rm R}$ is the Rayleigh drag frequency {(further details below)}. While the planet is subject to the tidal gravitational potential, $U_{\rm T}$, induced by the perturber, the equilibrium displacement corresponds to the tidal equipotential surface generated by $U_{\rm T}$, thus $\zeta_\mathrm{eq}=U_{\rm T}/g$.

We show here that it is straightforward to obtain a closed-form solution to these equations in the limit of a strongly viscous/highly frictional medium, which we adopt to describe tidal dynamics in the magma ocean. We characterize this regime by the condition $2\Omega\cos\theta\ll\sigma_{\rm R}$, or equivalently, a high Ekman number \citep[e.g.,][]{vallis2017atmospheric}{}{}. In other words, the Coriolis term in the momentum equation in this regime is relatively negligible and thus ignored as the dissipative term dominates.

{To this end, we emphasize that $\sigma_{\rm R}$ characterizes the damping timescale of the fluid tidal response by dissipative mechanisms, mainly how rotational and inertial fluid effects are subdued by these mechanisms. This frequency, albeit  appearing here as a frictional drag frequency, should nonetheless be considered as an \textit{effective} free parameter of energy dissipation. Namely, one might naturally attempt to relate this dissipative timescale to the material  viscosity in the magma ocean for a better physical interpretation of the dynamics at play. However, dissipation in the magma ocean is not restricted to viscous resistance against tidal flexure, but rather involves additional mechanisms, including: boundary layer friction; pressure (or form) drag, arising from local pressure variations when fluid lava flows against suspended solid inclusions; and Darcy dissipation when the magma ocean lives close to the rheological transition. For this reason, an inferred value of viscosity from $\sigma_{\rm R}$ can only serve as an effective viscosity which results from a non-trivial interplay between these different dissipative mechanisms. Therefore, in the results that follow, we treat $\sigma_{\rm R}$ as a free parameter that covers a wide range of possible values, ensuring, however, that over this whole range, the Coriolis term is negligible when compared to the dissipative term, and it is thus ignored.}

Under the said limit, we transform the governing system of \eq{momentum_continuity} to the Fourier domain using the tidal forcing frequency $\sigma$, and we expand the tidal quantities in spherical harmonics (Appendix \ref{App_Legendre_Functions}) of degree $n$ and order $m$ to obtain:
\begin{subequations}
\label{momentum_continuity_friction2}
\begin{align}\label{momentum_friction2} 
    &\Vec{u}_n = \frac{\grad{(U^{\rm T}_n-g\zeta_n)}}{\sigma_{\rm R}+i\sigma},\\
    &\grad\cdot\Vec{u}_n = -\frac{i\sigma\zeta_n}{H}. \label{continuity_friction2}
\end{align}
\end{subequations}
{Hereafter, we drop the longitudinal order $m$ from the harmonic components of the tidal variables, emphasizing that it is maintained in the tidal solution through the exponential complex term, $e^{im\lambda}$, of the spherical harmonics (Appendix \ref{App_Legendre_Functions}).} Taking the divergence of \eq{momentum_friction2} and equating it with \eq{continuity_friction2} we obtain a solution for the tidal amplitude $\zeta_n$ in the form
\begin{equation}\label{aux2}
        \frac{g\zeta_n}{U_n^{\rm T}} = -\frac{\bar\sigma_{n}^2}{\sigma\Tilde{\sigma}-\bar\sigma_{n}^2}.
\end{equation}
In this equation, we have defined the characteristic frequency $\bar\sigma_{n} = \sqrt{\mu_n gH}/R_{\rm p}$, where $\mu_n=n(n+1)$ is the eigenvalue of the $n$-th degree spherical harmonic, and 
\begin{equation}\label{sigma_tilde_def}
\Tilde{\sigma}=\sigma- i\sigma_{\rm R}.
\end{equation}
In an isolated tidally forced magma ocean, that is, ignoring the tidal response of the rest of the planet, the tidal displacement of \eq{aux2} generates a distortion tidal potential $U_n^{\rm D}$. In the classical LTEs, where Coriolis effects are included, mass redistribution and tidal energy distortion are scattered among different harmonic components, that is, different wavenumbers. Consequently, the harmonic distribution of the response does not necessarily entail a one-to-one correspondence with that of the tidal forcing, but rather a complex overlap via the so-called Hough functions \citep[e.g.,][]{hough,chapman1969atmospheric}{}{}, which can be described as the set of spherical harmonics distorted by rotation.  By virtue of the absence of Coriolis effects in our case, the distortion potential can be fully described by spherical harmonics, in contrast to the general case of fluid tides, where the Hough functions are adopted. The distortion tidal potential is given by \citep[][Eq. 2.1.25]{kaula1969introduction}
\begin{equation}\label{eq_potential_stress}
    U_n^{\rm D} = g\varrho_n\zeta_n,
\end{equation}
where $\varrho_n={3}/(2n+1)(\rho_{\rm f}/\rho_{\rm b})$ is the degree-$n$ density ratio between the fluid and the bulk part of the planet, also called the self-attraction term. The frequency-dependent, complex tidal Love number, which quantifies the intrinsic response of the magma ocean under tidal stresses, is defined as the ratio between the distortion tidal potential and that of the forcing, namely:
\begin{equation}\label{oceanic_LN}
k_n^\mathcal{O,\sigma} = \frac{U_n^{\rm D}}{U_n^{\rm T}} =-\frac{\varrho_n\bar\sigma_{ {n}}^2}{\sigma\Tilde{\sigma}-\bar\sigma_{ {n}}^2}.
\end{equation}
The long-term effects of magma tides on the planetary evolution scale are quantified by the tidal torque and the rate of tidal dissipation. Both these quantities are related to the imaginary part of the Love number in \eq{oceanic_LN} which is readily obtained, using \eq{sigma_tilde_def}, as
\begin{equation}\label{aux4}
    \mathfrak{Im} \left\{k_n^{\mathcal{O},\sigma}\right\}=-\frac{\varrho_n\sigma\sigma_{\rm R}\bar\sigma_{ {n}}^2}{(\sigma^2-\bar\sigma_{ {n}}^2)^2 + \sigma^2\sigma_{\rm R}^2} . 
\end{equation}
Hereafter, we use $\mathfrak{Im}\{x\}$ and $\mathfrak{Re}\{x\}$ to denote the imaginary and real parts of a complex quantity $x$. To further characterize the tidal response of the magma ocean, we take the derivative of \eq{aux4} to quantify the position and the amplitude of its maximum. We  have
\begin{equation}
   \frac{d}{d\sigma}\mathfrak{Im} \left\{k_n^{\mathcal{O},\sigma}\right\}= \varrho_n\sigma_{\rm R}\bar\sigma_{n}^2 \left\{ \frac{3\sigma^4 - \sigma^2(2\bar\sigma_{n}^2-\sigma_{\rm R}^2) - \bar\sigma_{n}^4}{\left[(\sigma^2-\bar\sigma_{n}^2)^2 + \sigma^2\sigma_{\rm R}^2\right]^2} \right\},
\end{equation}
for which the roots are 
\begin{equation}
    \sigma_{\rm p} = \pm \sqrt{\frac{1}{6}\left[ 2\bar\sigma_{n}^2  -\sigma_{\rm R}^2 \mp \sqrt{(2\bar\sigma_{n}^2  -\sigma_{\rm R}^2)^2 + 12\bar\sigma_{n}^4}\right]},
\end{equation}
with the plus sign inside the square root giving a real frequency. The limit of strong viscous damping and drag ($\sigma_{\rm R}\gg 2\Omega)$ within a thin magma shell ($H\ll R_{\rm p})$ justifies rearranging the latter expression using powers of the ratio $\bar\sigma_n/\sigma_{\rm R}$ and ignoring the high order terms. Doing so, the frequency of maximum tidal response simplifies to
\begin{equation}\label{peak_frequency}
      \sigma_{\rm p} \simeq \frac{\bar\sigma_{n}^2}{\sigma_{\rm R}}.
\end{equation}
\begin{figure}[t]
\includegraphics[width=.45\textwidth]{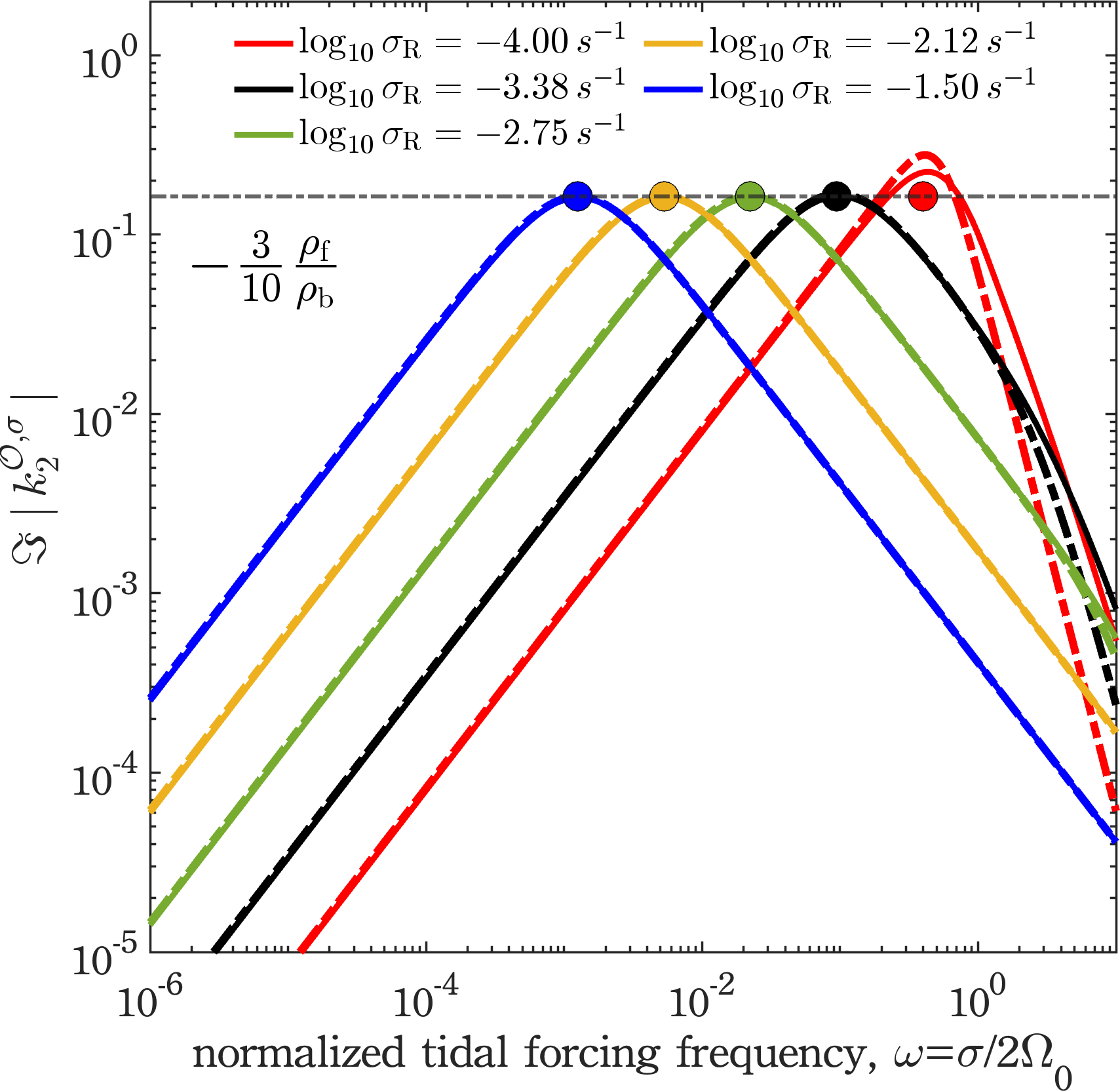}
\caption{The spectrum of the tidal response of an isolated strongly viscous/highly frictional medium representing a magma ocean. {The imaginary part of the quadrupolar ($n=2$) Love number ($k_2^{\mathcal{O},\sigma})$ is plotted as a function of the tidal forcing frequency, $\sigma=2(\Omega-n_{\rm orb})$, normalized by the present spin rate of the Earth ($\omega=\sigma/2\Omega_0$), for different dissipation frequencies, $\sigma_{\rm R}$. We fix $n_{\rm orb}$ to the present lunar orbital frequency, as an arbitrary forcing companion, and vary $\Omega$.} The solid-line spectra are plotted using the solution of the full LTEs, (i.e. allowing for Coriolis effects) following the formalism in \citet[][]{farhat2022resonant}{}{}, while dashed-line spectra follow the analytical expression of \eq{aux4}. {The colored dots mark the positions of the peaks using the analytical estimates of \eq{peak_frequency}, with amplitudes living on the dashed horizontal line and corresponding to the value obtained from \eq{peak_amp} with $n=2$}. We adopt $R_{\rm p}=6371$~km, $M_{\rm p}=5.97\times10^{24}$~kg, $H=50$ km, $\rho_{\rm f}=3000$~kg~m$^{-3}$.}
\label{Fig_love_number_magma}
\end{figure}
Substituting the solution of \eq{peak_frequency} in \eq{aux4}, we  retrieve the value of the imaginary part of the Love number at this peak, which characterizes the maximum value of the tidal response in the studied regime, as
\begin{equation}\label{peak_amp}
      \mathfrak{Im} \left\{k_n^{\mathcal{O},\sigma}\right\}\Big|_{\sigma=\sigma_{\rm p}} \simeq  -\frac{\varrho_n }{2}.
\end{equation}
Therefore, the tidal response of a forced magma ocean peaks at half the self-attraction between the ocean and the solid  bulk of the planet. We remark here that we have ignored in this section the direct tidal distortion of the solid part itself, which we have recovered in Section \ref{Section_Tides_Solid}, though it will be coupled to the fluid response in the next Section.

In Fig. \ref{Fig_love_number_magma} we show the frequency spectrum of the tidal response, quantified in terms of the imaginary part of the Love number, for different values of the damping frequency $\sigma_{\rm R}$. The spectra on solid curves are plotted using the full solution of the LTEs, allowing for Coriolis effects, using sets of Hough functions \citep[e.g.,][]{auclair2018oceanic,farhat2022resonant}{}{}, while the dashed curves correspond to \eq{aux4}. The colored circles mark our analytical estimates for the positions (Eq.~\ref{peak_frequency}) and amplitudes (Eq. ~\ref{peak_amp}) of the peaks in the magma flow limit. Our analytical estimates are thus in good agreement with the full solution of the LTEs when strong damping frequencies are considered \citep[typically,  $\sigma_{\rm R}$ with values of $10^{-5}$ s$^{-1}$ and smaller are considered when modeling water oceans; e.g., ][]{webb1982tides,matsuyama2014tidal,farhat2022resonant}{}{}. Evidently, however, the two solutions diverge when $\sigma_{\rm R}$ decreases and  approaches the typical rotational velocity of the planet, i.e. when Coriolis effects are no longer expected to be negligible.

This evident and robust broad peak of enhanced tidal response is the characteristic signature of the magma ocean, and we shall focus on it in the rest of this paper. As it cannot be captured when magma tides are modeled in the conventional solid deformation formalism, it highlights the necessity of employing the LTEs. It is important to emphasize, however, that this peak is not associated with a resonant mode of fluid tides, akin to those retrieved in water oceans \citep[e.g.,][]{tyler2011tidal,matsuyama2014tidal,kamata2015tidal}. In fact, gravity modes responsible for these resonances are excited at higher rotation rates, i.e. when Coriolis effects are present,  and they require weak damping to survive \citep[e.g.,][]{auclair2018oceanic,farhat2022resonant,auclair2023can}{}{}, both features being absent in our overdamped regime. The peak in the response can be simply understood by the interplay between two rates: the typical velocity of long-wavelength gravity waves propagation within a thin shell, and the rate at which these waves dissipate energy, as can be seen by \eq{peak_frequency}. Enhanced dissipation, for a fixed rate of horizontal momentum propagation, typically tends to attenuate the response of the tidally forced layer. In this regime, however, enhanced dissipation also affects the propagation rate, allowing for an interplay of competing effects with two rates and inducing the peak in the tidal response.

\subsection{The Tidal Response of a (sub-)Surface Magma Ocean}\label{Section_magma_ocean_subsurface}
Having estimated the tidal response of an isolated magma layer, we now study the general case of a magma ocean that lives on a planet, on top of a solid mantle, and possibly underlying a crust. The general setting of a subsurface magma ocean can be easily reduced, as we elaborate towards the end of this section, to a surface magma ocean in the absence of a crust. {The two described configurations are illustrated in the schematic of Figure \ref{magma_sketch}}. To recover the tidal response of the whole planet we combine the solution of the LTEs obtained for an isolated magma ocean with the tidal response of the solid mantle and that of the crust, allowing for coupling and self-attraction effects between the three layers. Namely, presuming that the crust is thin on the planetary scale ($d/R_{\rm p}{\sim}5 {-} 10 \%$), and that crustal deformation occurs over wavelengths that are much longer than its thickness (quadrupolar tidal deformations have a wavelength of half the planet's circumference), we adopt the membrane theory of shells to resolve the tidal response of the crust \citep[e.g.,][]{beuthe2016crustal,matsuyama2018ocean,beuthe2019enceladus,rovira2023thin}{}{}. The theory, in the elastic limit, provides analytical solutions for tidal stresses \citep[e.g,][]{meinesz1947shear,greenberg1998tectonic}{}{}, and has been generalized to allow for crustal viscoelasticity \citep[][]{beuthe2015tides}.

{In the presence of a crust (Figure \ref{magma_sketch}, model B)}, the tidal displacement of the magma ocean is defined by the difference between the displacements of the ocean-crust interface, $\zeta_n^{\rm \uparrow}$, and the ocean-solid mantle interface,  $\zeta_n^{\rm \downarrow}$; namely:
\begin{equation}\label{radial_displacement}
    \zeta_n = \zeta_n^{\rm \uparrow}-\zeta_n^{\rm \downarrow}.
\end{equation}
The distortion potential of the planet (Eq. \ref{eq_potential_stress}) now reads \citep[e.g.,][]{beuthe2016crustal}{}{}:
\begin{equation} \label{distortion_potential_1}
    U_n^{\rm D} = k_n^{\rm T}U_n^{\rm T} + (1+k_n^{\rm L})U_n^{\rm L} + k_n^{\rm P} U_n^{\rm P},
\end{equation}
where $U_n^{\rm L}$ and $U_n^{\rm P}$ are respectively the mass-load and pressure potentials, while $k_n^{\rm P}$ is the pressure Love number. As such, the first term delivers the distortion of the solid mantle, while the second codes for the loading and self-attraction between the solid mantle and the magma ocean. While the magma-crust interface is treated as a free-slip solid-liquid boundary, there is no direct gravitational self-attraction between the two layers, leaving us with the pressure potential only, present in the third term. In analogy to \eq{eq_potential_stress}, the loading and pressure potentials are expressed as \citep[][]{kaula1969introduction}{}{}
\begin{equation}
    U_n^{\rm L} = \frac{3}{2n+1}\frac{g \mathcal{S}_n}{\rho_{\rm b}}, \,\,\, U_n^{\rm P} = \frac{3}{2n+1}\frac{\mathcal{Q}_n}{\rho_{\rm b}},
\end{equation}
where $\mathcal{S}_n$ is the surface density of the mass load, and $\mathcal{Q}_n$ is the surface pressure. The surface density of the oceanic tidal loading is expressed in terms of the oceanic displacement to obtain
\begin{equation}
     U_n^{\rm L} = g \varrho_n \zeta_n,
\end{equation}
while the loading pressure is related to the radial deformation of the crust via Hooke's law \citep[][]{beuthe2016crustal}{}{}:
\begin{equation}\label{loading_pressure}
    \mathcal{Q}_n = \rho_{\rm f} g \Lambda_n^{\rm M}\zeta_n^{\rm \uparrow}.
\end{equation}
In this expression, the membrane spring constant $\Lambda_n^{\rm M}$, a function of the crustal thickness and rigidity, is a non-dimensional parameter that characterizes crustal resistance against extension. \citet[][]{beuthe2016crustal}{}{} provides its explicit form for a crust characterized by a Maxwell rheology, allowing for radial variations in the physical properties. Here we assume that the crust has a constant thickness and uniform shear modulus and viscosity, and we adopt an Andrade rheology, similar to that of the solid mantle. We further assume, for simplicity, that the crust and the underlying magma layer have the same density, $\rho_{\rm f}$. To obtain the explicit form of the membrane constant for an Andrade rheology, we start with the equations governing crustal flexure in the membrane and thin shell limits\footnote{We obtain these equations directly from Eqs. (86-87) of \citet[][]{beuthe2008thin}{}{}, by imposing the membrane approximation (vanishing bending rigidity, i.e. setting their $D$ to $0$), the thin shell approximation (setting their $\chi$, a function of the ratio 
 of crustal thickness to $R_{\rm p}$, to $\infty$), and switching the sign of the pressure load to correct for the loading direction ($\mathcal{Q}_n\rightarrow-\mathcal{Q}_n$).}\citep[][]{beuthe2008thin}:
\begin{subequations}
\label{flexure_eqns}
\begin{align}\label{flexure1}
&\Delta^\prime F_n = R_{\rm p} \mathcal{Q}_n - 2\Psi_n,\\
&(\Delta^\prime -1 -\nu)\Delta^\prime F_n= \frac{d}{R_{\rm p}}E \Delta^\prime w_n - (1-\nu)\Delta^\prime \Psi_n. \label{flexure2}
\end{align}
\end{subequations}
These equations relate the stress function $F_n$ and the vertical pressure load $\mathcal{Q}_n$ to the crustal deflection $w_n$ (which in our case corresponds to the ocean's top radial displacement, $\zeta_n^{\rm \uparrow})$ and the tangential loading potential $\Psi_n$. The operator $\Delta^\prime=\Delta+2$, where $\Delta$ is the spherical Laplacian, {the restriction of the Laplacian in three dimensions to the surface of the unit sphere (thus $\Delta^\prime$ admits the eigenvalues $\lambda_n=-(n-1)(n+2))$}, $d$ is the crustal thickness, while $E$ and $\nu$ are the Young modulus and the Poisson ratio of the crust, respectively, {the latter being the ratio of strains (lateral to radial) as the crust is stressed radially.} $E$ and $\nu$ deliver the viscoelasticity of the crust and encode its assumed rheology.  

Since we are interested with the setting of a magma ocean underneath the crust, a free slip condition at their interface can be assumed whereby shear forces exerted by the ocean on the crust vanish, and consequently the tangential potential $\Psi_n$. As such, one can combine the two equations (\ref{flexure1}-\ref{flexure2}) into a single equation:
\begin{equation}
    R_{\rm p}^2(\Delta^\prime -1 -\nu)  \mathcal{Q}_n = 2 d \mu(1+\nu)\Delta^\prime \zeta_n^{\rm \uparrow},
\end{equation}
where we have used the equivalence of the shear modulus $\mu  = E/(2+2\nu)$. Using the definition of the loading pressure in \eq{loading_pressure}, the membrane constant expression is then readily obtained in the form:
\begin{equation}\label{membrane_constant}
    \Lambda_n^{\rm M} = \frac{2d}{R_{\rm p}}\frac{\mu}{g\rho_{\rm f}R_{\rm p}}\frac{\lambda_n(1+\nu)}{(\lambda_n + 1 +\nu)}.
\end{equation}
We provide the explicit expressions of the shear modulus and the Poisson ratio for an Andrade rheology in Appendix \ref{Appendix_Crust}. With these definitions, one can now write the distortion potential of \eq{distortion_potential_1} in the form:
\begin{equation}\label{distortion_potential_2}
     U_n^{\rm D} = k_n^{\rm T}U_n^{\rm T} + g\varrho_n\left[(1+k_n^{\rm L})\zeta_n - h_n^{\rm T} \Lambda_n^{\rm M}\zeta_n^{\rm \uparrow}\right], 
\end{equation}
where we have used the equivalence of the pressure Love numbers, $h_n^{\rm P}= h_n^{\rm L}-h_n^{\rm T}$ and $k_n^{\rm P}=-h_n^{\rm T}$ \citep[e.g.,][]{hinderer1986resonance}{}{}. Analogously, the magma ocean-mantle interface displacement $\zeta_n^{\rm \downarrow}$ is expressed as
\begin{equation}\label{bottom_displacement}
   \zeta_n^{\rm \downarrow} = h_n^{\rm T} U_n^{\rm T}/g  +\varrho_n \left[ h_n^{\rm L} \zeta_n + (h_n^{\rm L}-h_n^{\rm T})\Lambda_n^{\rm M}\zeta_n^{\rm \uparrow} \right].
\end{equation}
Using \eq{bottom_displacement} with \eq{radial_displacement}, one obtains for the displacement of the magma ocean-crust interface:
\begin{equation}
    \zeta_n^{\rm \uparrow} = \frac{h_n^{\rm T}U_n^{\rm T}/g + (1+\varrho_n h_n^{\rm L})\zeta_n}{1-\varrho_n(h_n^{\rm L}-h_n^{\rm T})\Lambda_n^{\rm M}}.
\end{equation}
\begin{figure*}[ht]
\includegraphics[width=\textwidth]{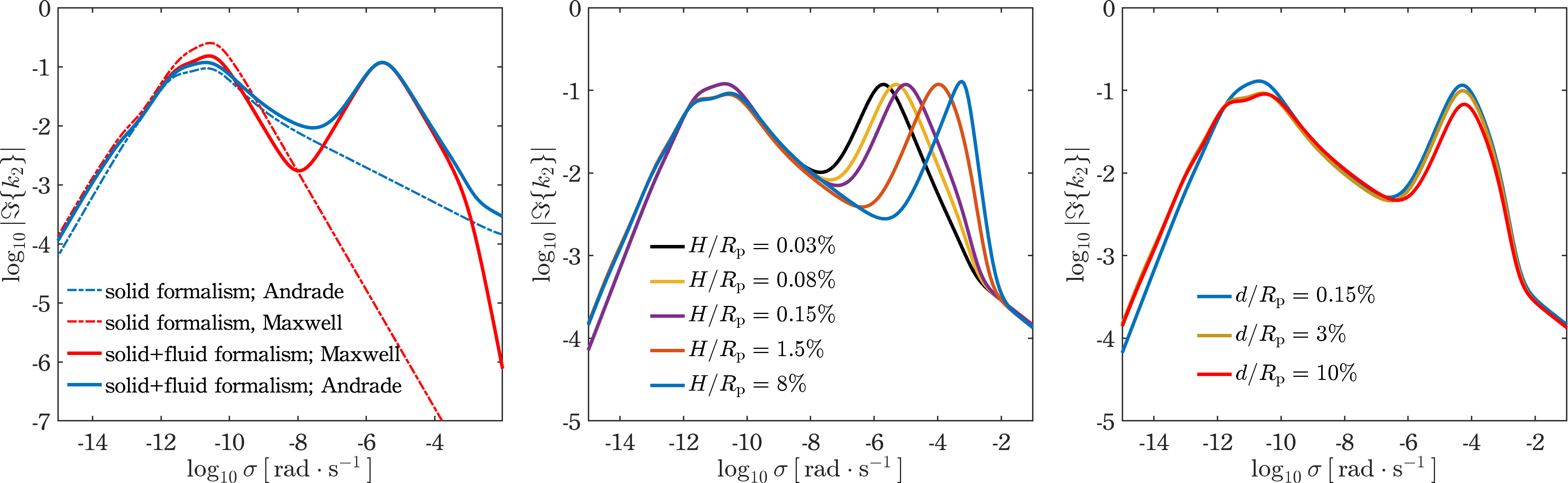}
\caption{The frequency spectrum of the imaginary component of the planetary quadrupolar Love number under different variations. \textit{Left:} The planetary tidal response is computed with the solid mantle characterized by an Andrade (blue) or Maxwell (red) rheology, while the magma ocean responds rheologically as a solid (dashed) or a fluid (continuous). \textit{Middle:} We vary the thickness of the magma ocean, assuming it resides on the surface and that it responds rheologically as a fluid, while the solid counterpart follows an Andrade rheology. \textit{Right:} We fix the thickness of the magma ocean to $H/R_{\rm p}=1.5\%$, and vary the thickness of the solid crust at the surface. Similar to Figure~\ref{Fig_love_number_magma}, $R_{\rm p}=6371$~km, $M_{\rm p}=5.97\times10^{24}$~kg, $\rho_{\rm f}=3000$~kg~m$^{-3}$, with $\sigma_{\rm R}= 10^{-3}$~s$^{-1}$.}
\label{Fig_Love_numbers}
\end{figure*}
What remains to fully define the distortion potential (Eq. \ref{distortion_potential_2}) is therefore the displacement solution $\zeta_n$. In \eq{aux2}, we provided the solution for an isolated magma ocean. For a subsurface magma ocean, however, the forcing term in the momentum equation of the LTEs is given by $g\beta_n\zeta_n - \upsilon_n U_n^{\rm T}$ \citep[][]{beuthe2016crustal}{}{}, where the coefficients $\beta_n$ and $\upsilon_n$ are defined as:
\begin{subequations}
\label{crustal_coeffs}
\begin{align}\label{beta_coeff}
&\beta_n = 1-\varrho_b\gamma_n^{\rm L}  + \Lambda_n^{\rm M} + \delta\Lambda_n^{\rm M},\\
&\upsilon_n =\gamma_n^{\rm T} + \delta\gamma_n^{\rm T}.\label{upsilon_coeff}
\end{align}
\end{subequations}
In these expressions, the loading and tidal solid deformation factors, $\gamma_n^{\rm L}$ and $\gamma_n^{\rm T}$ respectively, are given by \citep[e.g.,][]{matsuyama2014tidal,auclair2023can}{}{}:
\begin{equation}
   \gamma_n^{\rm L} = 1+k_n^{\rm L} -h_n^{\rm L} ; \,\,\,\, \gamma_n^{\rm T} = 1+k_n^{\rm T} -h_n^{\rm T},
\end{equation}
while the crust-mantle coupling coefficients read as \citep[][]{beuthe2016crustal}{}{}:
\begin{subequations}
\label{crustal_mantle_coeffs}
\begin{align}\label{del_lambda}
&\delta\Lambda_n^{\rm M} = -\left(1- \frac{(1+\varrho_n h_n^{\rm L})^2}{1+\varrho_n(h_n^{\rm T}-h_n^{\rm L})\Lambda_n^{\rm M}} \right)\Lambda_n^{\rm M},\\
&\delta\gamma_n^{\rm T} = -\left(1- \frac{1+\varrho_n h_n^{\rm L}}{1+\varrho_n(h_n^{\rm T}-h_n^{\rm L})\Lambda_n^{\rm M}}h_n^{\rm T} \right)\Lambda_n^{\rm M}. \label{del_gamma
}
\end{align}
\end{subequations}
The solid deformation coefficients thus tend toward unity in the limit of an infinitely rigid mantle ($\mu_{\rm E}\rightarrow\infty)$, while the crust-mantle coupling coefficients vanish in the limit of a fluid crust ($\mu_{\rm E}\rightarrow0)$ or a rigid mantle. With these definitions, the displacement solution of the magma ocean, given earlier by \eq{aux2}, now reads
\begin{equation}\label{aux3}
        \frac{g\zeta_n}{U_n^{\rm T}} = -\frac{\bar\sigma_{n}^2\upsilon_n}{\sigma\Tilde{\sigma}-\bar\sigma_{n}^2\beta_n}.
\end{equation}
We are now fully geared to define the self-consistent tidal response of the whole planet with a deformable membrane crust, a sub-surface magma ocean, and a viscoelastic mantle, using the frequency-dependent, dynamical Love number $k_n = U_n^{\rm D}/U_n^{\rm T}$, which gives:
\begin{align}\nonumber
    k_n &= k_n^{\rm T} - \frac{\varrho_n (h_n^{\rm T})^2 \Lambda_n^{\rm M}}{1-\varrho_n h_n^{\rm P}\Lambda_n^{\rm M}}-\left[1+k_n^{\rm L} - \frac{h_n^{\rm T}\Lambda_n^{\rm M}(1+\varrho_n h_n^{\rm L})}{1-\varrho_n h_n^{\rm P}\Lambda_n^{\rm M}}\right]\\
    &\times\varrho_n\frac{\bar\sigma_{n}^2\upsilon_n}{\sigma\Tilde{\sigma}-\bar\sigma_{n}^2\beta_n}. \label{eq_full_LN}
\end{align}
This Love number is the key takeaway from this formalism. If the magma ocean resides on the surface of the planet {(Figure \ref{magma_sketch}, model A)}, it is straightforward to recover the planet's tidal response by ignoring the terms that include the crustal membrane constant, $\Lambda_n^{\rm M}$ in \eq{eq_full_LN}. This will be the case of the studies we perform in Sections \ref{Section_Early_Earth_Moon} and \ref{Section_Close_In_Exo}.

In Figure \ref{Fig_Love_numbers} we plot the frequency spectrum of the quadrupolar component of this Love number allowing for several variations. We focus on the imaginary part of the Love number since $\mathfrak{Im}\{k_n\}$ quantifies tidal dissipation and drives the orbital and rotational evolution of the planet via the generated torque. In the left panel, we compute the tidal response of an Earth-like planet with a sub-crust magma ocean using two approaches: $\textit{i)}$ using \eq{eq_full_LN}, where the magma ocean is treated as a strongly viscous fluid and the solution of the LTEs in the high friction limit is adopted, and $\textit{ii)}$ using a fully solid formalism, where the magma ocean is treated as a low viscosity solid, as often assumed in earlier studies. We do so for two rheological descriptions of the solid part present in either model: a Maxwell and an Andrade rheologies. 

In both approaches, a peak of the tidal response is present at smaller forcing frequencies, one that is associated with a material resonance in the anelastic solid \citep[e.g.,][]{nowick2012anelastic}{}{}. In the fluid description, as we have structured the LTEs in the overdamped limit, we have suppressed all possible resonant excitations that are typically recovered in fluid tides. As such, similarities are  expected between the fluid and solid descriptions of magma tides. However, the  significantly different material properties between the two media favor larger tidal displacements in the stressed fluid, thus requiring, by virtue of the continuity equation in the fluid, stronger flow momentum within the thin layer to fill these displacements. Therefore, the fluid treatment of the magma ocean, compared to the solid one, delivers the additional peak of the tidal response recovered in Section \ref{Section_Magma_isolated}. This peak occurs toward higher tidal forcing frequencies and, for an Andrade rheology, it amplifies the tidal response of the planet relative {to the pure solid response by two orders of magnitude}. In contrast, for a Maxwell rheology, where the anelastic response of the solid decays much faster at higher frequencies than for an Andrade rheology, the relative amplification in the tidal response is that of several orders of magnitude. As for the peak associated with the anelastic solid response, while it is expected that the fluid treatment of the magma ocean would alter this peak by reducing the total volume of the viscoelastic solid, it is noteworthy that this effect is not the same for both solid rheologies. Namely, the amplitude of the anelastic peak is decreased (increased) by the fluid for a Maxwell (Andrade) rheology; an effect that is worthy of further investigation. 

In the middle panel of Figure \ref{Fig_Love_numbers}, we adopt the fluid treatment of magma tides and we explore the effect of the magma ocean thickness on the spectrum. As we showed in \eq{peak_amp}, the position of the peak of the fluid tidal response is a linear function of $H/R_{\rm p}$. As such, for a fixed dissipation timescale, increasing the thickness of the magma layer drags the peak toward higher forcing frequencies. Consequently, since the solid tidal response decays at higher frequencies, increasing $H$ brings about stronger tidal amplification relative to the pure solid treatment. This does not mean that increasing $H$ amplifies the absolute tidal dissipation due to the larger volume of fluid that is tidally stressed. On the contrary, for some forcing frequencies, thinner magma layers deform and dissipate energy more strongly than thicker ones. This is by virtue of the fact that dissipation is a strong function of the flow momentum in the magma layer, which in turn is inversely proportional to the layer's thickness. It is therefore noteworthy that the tidal signature of the magma ocean in the fluid regime is significant even for a very thin magma layer. In the last panel of Figure \ref{Fig_Love_numbers}, we show the effect of the crustal thickness, $d$, on the spectrum of the Love number. As thoroughly investigated by \citet[][]{beuthe2016crustal}{}{}, a viscoelastic crust damps the response of the underlying fluid; an effect that we capture in this panel, noting the logarithmic scale of the Love number.  While one might naturally expect that increasing the crustal thickness should substantially damp the barotropic response of the underlying fluid, as the latter would then be sandwiched in between two solid layers, the current formalism does not allow for this behavior since the crust is merely modeled as a membrane without a gravitational potential.

\subsection{The Resultant Tidal Dissipation}
Before embarking on the physical applications of the formalism, we finalize this section by computing the resultant tidal dissipation of a planet harboring a surface (or subsurface) magma ocean, and subject to the gravitational potential $U^{\rm T}$ of a perturber with mass $M^\prime$. The planet rotates with a velocity $\Omega$, while the planet-perturber binary is characterized by the orbital semi-major axis $a$, mean motion $n_{\rm orb}$, and eccentricity $e$. Our fiducial system is fully coplanar, that is, we set the planetary obliquity and mutual orbital inclination with the perturber to zero. By virtue of energy conservation, the tidal dissipative work of the whole planet, $\mathcal{P}_{\rm diss}$, is equal to the work done on the planet by the perturber's tidal forces, $\mathcal{P}_{\rm T}$. Per unit time, the latter is defined as \citep[e.g.,][]{ogilvie2013tides}{}{}:
\begin{equation}\label{dissp_1}
    \mathcal{P}_{\rm T} = \left\langle \int_{\mathcal{V}_{\rm p}}\rho \vec V\cdot \grad U^{\rm T} dV \right\rangle,
\end{equation}
where $\mathcal{V}_{\rm p}$ is any connected region that includes the planet but not the perturber, $\vec V$ is the velocity field of the tidal flow in the whole planet, $\rho$ is the local density, and $dV$ is an infinitesimal volume element. Integrating  \eq{dissp_1} by parts, and making use of mass conservation and Poisson's equation as detailed in \citet[][]{auclair2023can}{}{}, we obtain:
\begin{equation}\label{Eq_dissipation}
    \mathcal{P}_{\rm T} = \frac{R_{\rm p}}{8\pi G}\sum_{s=-\infty}^\infty\sum_{n=2}^\infty\sum_{m=0}^n{(2n+1)\sigma_{m}^s}{}|U_{n,m}^{\rm T; s}|^2 \,\mathfrak{Im}\left\{k_{n}^m\right\},
\end{equation}
where $G=6.67430\times10^{-11}{\rm m^3 \ kg^{-1} \ s^{-2} }$ is the gravitational constant \citep[][]{tiesinga2021codata}{}{}, $\sigma_m^s = m\Omega-s n_{\rm orb}$ is the tidal forcing frequency, and $U_{n,m}^{\rm T; s}$ is the $(n,m)-$harmonic and $s-$Fourier component of the gravitational potential $U^{\rm T}$; namely, in a coplanar setting \citep[e.g.,][]{kaula1969introduction}{}{}:
\begin{equation}
    U^{\rm T} = \mathfrak{Re}\left\{\sum_{s=-\infty}^\infty\sum_{n=2}^\infty\sum_{m=0}^n U_{n,m}^{\rm T; s} P_n^m \exp{i[\sigma_m^s t+m\lambda]}\right\},
\end{equation}
with $P_n^m$ denoting the set of normalized associated Legendre functions (Appendix \ref{App_Legendre_Functions}), and 
\begin{equation}
   U_{n,m}^{\rm T; s}  = \frac{GM^\prime}{a}\left(\frac{R_{\rm p}}{a}\right)^n A_{n,m,s}(e),
\end{equation}
where the dimensionless functions $A_{m,n,s}(e)$ are defined as:
\begin{align}\nonumber
    A_{m,n,s}(e) &= (2-\delta_{m,0}\delta_{s,0})(1-\delta_{m,0}\delta_{s<0})\\
    &\times \sqrt{\frac{2(n-m)!}{(2n+1)(n+m)!}}P_n^m(0)X_s^{-(n+1),m}(e).
\end{align}
\begin{figure}[t]
\includegraphics[width=.47\textwidth]{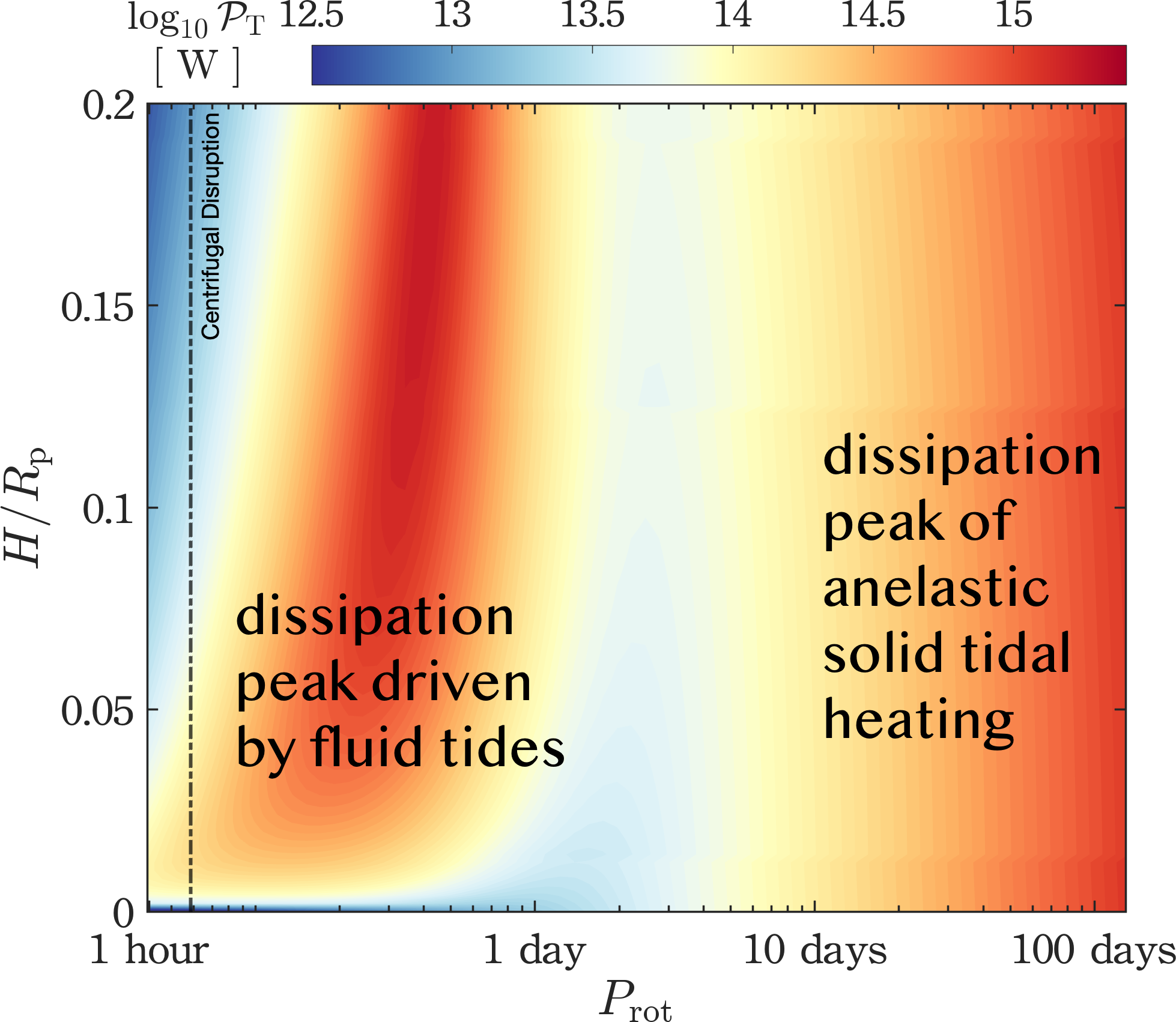}
\caption{Power dissipated by an Earth-like planet orbiting around a $1 M_{\odot}$ star at 0.7 AU with eccentricity $e=0.5$. On the x-axis, we vary the rotational period of the planet to map variations in the tidal forcing frequency at fixed orbital distance. The planet harbors a surface magma ocean, of varying thickness, for which the tidal response is computed via the quadrupolar Love number of \eq{eq_full_LN}. We fix the dissipative frequency of the ocean to $\sigma_{\rm R}=10^{-3}$~s$^{-1}$.}
\label{Fig_PT_contour}
\end{figure}
In the above expression, the Kronecker delta function 
$\delta_{i,j}=1$ if $i=j$ and 0 otherwise, $P_n^m$ are the normalized associated Legendre functions \citep[e.g.,][]{abramowitz1972handbook}{}{}, and $X_s^{n,m}(e)$ are the Hansen eccentricity functions \citep[e.g.,][]{hughes1981computation,laskar2005note}{}{}. Evidently, a choice has to be made on the truncation order of the harmonic series and the Fourier expansion. In the limit of $R_{\rm p}\ll a$, it suffices to only consider the quadrupolar harmonic ($n=2)$. Moreover, as we are considering a coplanar geometry, terms with $m=1$ associated with obliquity tides vanish. Finally, the truncation order of the Hansen series is a function of the orbital eccentricity. Namely, the more eccentric the orbit is the larger the truncation order should be for the series to converge.  In Figure \ref{Fig_PT_contour}, we show a example of the parametric dependence of tidal heating of an Earth-like planet with a surface magma ocean (i.e., without a solid crust at the surface) on the rotational period of the planet, and the magma ocean thickness. As opposed to the standard response of the solid interior  of the planet, the presence of the magma ocean interrupts the smooth decay of energy dissipation as a function of the increasing rotational velocity by inducing a parametric region of enhanced dissipation towards the high rotational velocity end of the parameter space. This region maps the peak of the tidal response that we have recovered in the previous sections and shown in Figures \ref{Fig_love_number_magma} and \ref{Fig_Love_numbers}. 

\section{The Early Earth-Moon System}\label{Section_Early_Earth_Moon}
As a first application of the formalism, we consider the tidal evolution of the early Earth-Moon system. It is suggested that the energy released during the Moon-forming giant impact have formed a magma
ocean on Earth \citep[e.g.,][]{tonks1993magma,canup2012forming,nakajima2015melting}{}{}. The extent of melting, and consequently the depth of the magma ocean, and the degree of mixing and equilibration between the two impact bodies depend on
the impactor's size, velocity, and impact angle \citep[e.g.,][]{dahl2010turbulent,nakajima2021scaling}{}{}. Our purpose in this section is to isolate and study the effect of tidal dissipation in this terrestrial magma ocean using the developed formalism. 


{In the aftermath of the impact and chemical equilibration,  modeling lava tides is important to understand the early orbital dynamics of the Moon and identify what regime of Lunar orbital recession they would drive. The lifetime of this post-impact magma ocean is controlled by an interplay between multiple contributions, including: the ocean's composition and  rheological behavior, the interior convective regimes at play,  and whether the early atmosphere was rich in greenhouse gases for thermal blanketing to be effective \citep[e.g.,][]{solomatov2007magma,elkins2008linked,zahnle2015tethered,monteux2016cooling}{}{}. The latter effect, in particular, was studied by \citet{zahnle2015tethered}, who assumed that the early atmosphere contained significant amounts of water vapor, thereby slowing down the cooling of the magma ocean and allowing it to persist in a fluid rheology for ${\sim}5\,{\rm Myrs}$. A ``hard magma ocean" -- in which fluid lava percolates through an interconnected  solid matrix -- then follows until complete solidification  occurs ${\sim}100\,{\rm Myrs}$. The Moon, according to \citet[][]{zahnle2015tethered} must have expanded in orbit from few Earth radii ($R_{\rm E})$ to beyond 40 $R_{\rm E}$ upon the Earth's complete solidification. Such a scenario, however, is based on the assumption that the Earth's mantle was described by a uniform adiabatic structure, that is, the mantle's solidus, liquidus, and adiabats are parallel at all depths\footnote{The recent work of \citet[][]{korenaga2023rapid} described the mantle structure adopted by \citet{zahnle2015tethered} as isothermal. The latter, however, did not assume an isothermal mantle, but rather a uniform adiabatic structure described by a single potential temperature value.}. More recently, \citet[][]{korenaga2023rapid}{}{}, while ignoring the effect of atmospheric thermal blanketing, corrected for the proper adiabatic profile of the mantle, and concluded that the lifetime of the fluid magma ocean is on the order of $10^4\,{\rm yrs}$.} Consequently, Lunar recession during the fluid magma ocean phase is limited and delivers a  Moon that would still be within 10 $R_{\rm E}$
upon the complete solidification of the Earth. 

Common among these and similar studies, however, is the treatment of magma tides in the framework of solid tides, even when the melt fraction is well above its critical value. With this limitation in mind, our motivation in this section is two-fold: first, to investigate whether enhanced tidal dissipation associated with the fluid behavior of the magma ocean, which we have uncovered in the previous sections, can affect the timescale of the ocean's survivability; and second, to explore the effect of this enhanced dissipation on early Lunar recession. As such, and to better view our results in context of earlier studies, we consider a simplified scenario where we only focus on terrestrial tides that are driven by an infant Moon, which had formed just outside the Earth's Roche limit, on a circular and coplanar orbit. Several modeling choices will also be consistent with earlier studies, especially with the recent work of \citet[]{korenaga2023rapid}{}{}. {Namely, in order to isolate the effect of fluid tides within the magma ocean, we also simplify the model and assume that the early atmosphere was dry and lacked significant amounts of greenhouse gases, besides CO$_2$, rendering its thermal blanketing effect negligible \citep[][]{miyazaki2019timescalea}.} We delegate the discussion on how to improve some of these choices to the Discussions section. 

\subsection{Mantle adiabats and melting curves}
To construct our coupled thermal-orbital evolution system, we structure the phase diagram of the Earth's mantle by adopting parameterized cubic spline functions for the solidus, $T^{s}(z)$, the liquidus, $T^{l}(z)$, and the temperature profile $T^{40}(z)$ of the critical melt fraction $(F_{\rm m, c}=40\%)$  \citep[][]{zhang1994melting,fiquet2010melting,miyazaki2019timescalea}{}{}. For temperatures above (below) $T^{l}$ ($T^{s}$), the silicate mantle is fully molten (solid), while $T^{40}$ defines the rheological transition from a deformable solid matrix that is penetrated by a network of percolating melt to a fluid magma layer with suspended solids \citep[e.g.,][]{abe1997thermal}{}{}. We define these profiles as \citep[see Eqs.(1-4) of][]{korenaga2023rapid}{}{}:
\begin{equation}
    T^{s}(z) = \sum_{i=1}^3\left[H(z-z_i)-H(z-z_{i+1})\right]\times S_i^{s},
\end{equation}
\begin{equation}
    T^{l}(z) = \sum_{i=1}^3\left[H(z-z_i)-H(z-z_{i+1})\right]\times S_i^{l},
\end{equation}
\begin{equation}
    T^{40}(z) =  T^{s}(z) + f_{40}(z)\times\left[T^{l}(z) -T^{s}(z)\right],
\end{equation}
where $z$ is the planetary depth in km, $H$ is the Heaviside step function $[H(x) = 0$ for $x < 0$ and $H(x) = 1$ for $x\geq0]$, $z_i$ is the depth of the $i$-th knot in km, with four knots chosen at 0, 410, 670, and 2900 km, à la \citet[][]{korenaga2023rapid}{}{}. The cubic spline $S_i^{ s/l}$ is defined as:
\begin{align}\nonumber
 S_i^{ s/l} =& \frac{b_{i+1}^{ s/l}(z-z_i)^3}{6h_i} + \frac{b_{i}^{s/l}(z_{i+1}-z)^3}{6h_i}   \\\nonumber
&+\left[\frac{T^{s/l}_{i+1}}{h_i} - \frac{b^{ s/l}_{i+1}h_i}{6}\right](z-z_i) \\
&+\left[\frac{T^{s/l}_{i}}{h_i} - \frac{b^{s/l}_{i}h_i}{6}\right](z_{i+1}-z), 
\end{align}
where $h_i=z_{i+1}-z_i$, $T_i^{s/l}$ and $b_i^{s/l}$ are the values of the function and their second derivatives at the $i$-th knots, namely: $T_i^{s} = \{1273, 2323, 2473, 3985\}$~K; $T_i^{l}= \{1973, 2423, 2723, 5375 \}$~K; $b_i^{s}=-10^{-3}\times\{1, 7, 1, -0.3 \}$; $b_i^{l}=-10^{-3}\times\{1, 7, 2.5, -0.5 \}$. Finally, the function $f_{40}(z)$ is a linear interpolation between 0.4 at 0 km, 0.4 at 410 km,
0.6 at 670 km, and 0.845 at 2900 km. 

\begin{figure}[t!]
\includegraphics[width=.47\textwidth]{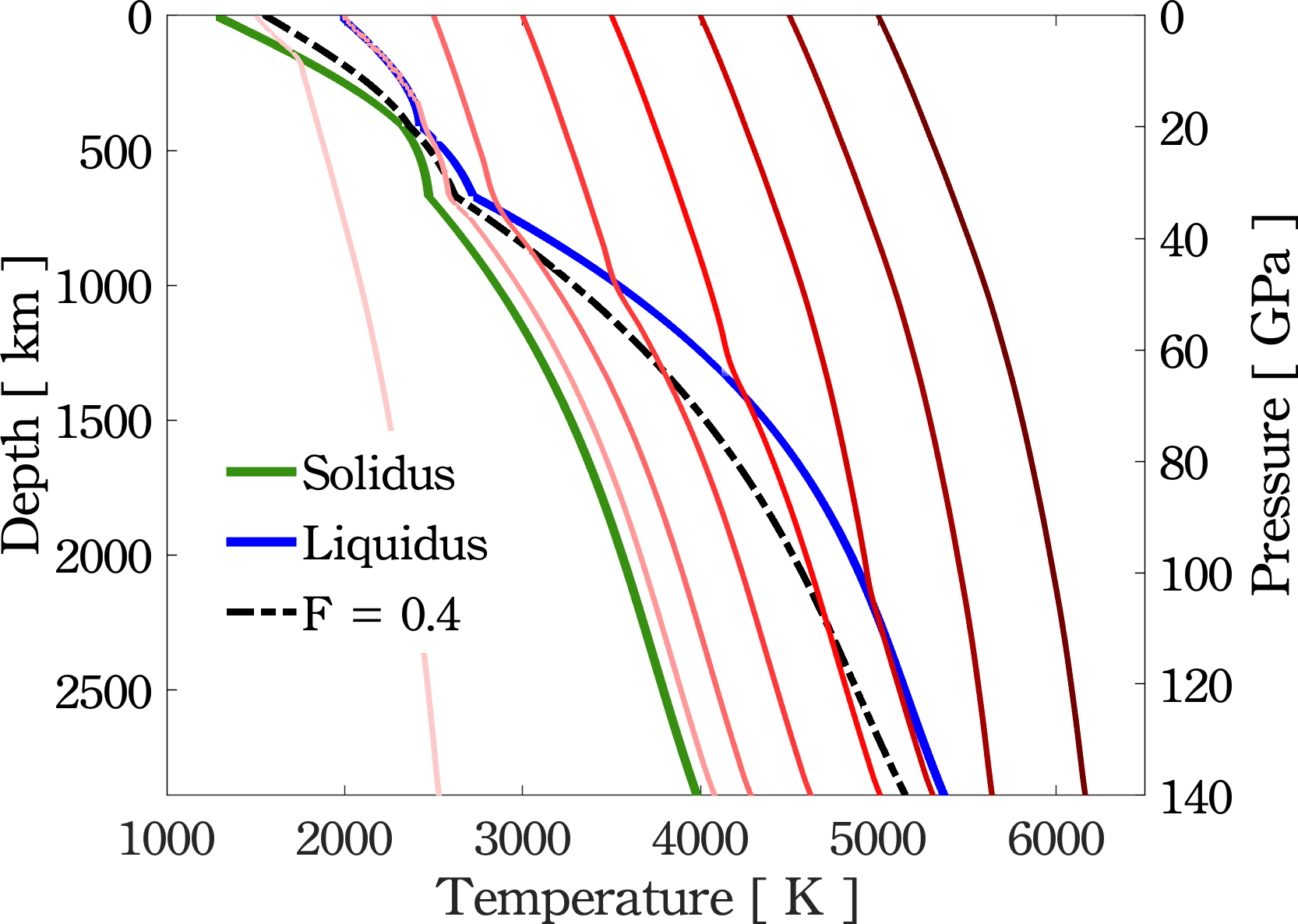}
\caption{The adopted mantle phase diagram of the Earth. Adiabatic temperature profiles with potential temperatures ranging from 1500 K (light pink) to 5000 K (dark red) in steps of 500 K are shown, along with the temperature profiles of the solidus (solid green), the liquidus (dark blue) and the critical melt fraction, $F_{\rm m,c}=40\%$ (dashed), which marks the rheological at each mantle depth. }
\label{Fig_Mantle_adiabats}
\end{figure}

Throughout the magma ocean's lifetime, the thermal structure of the mantle is assumed to be adiabatic. The choice is justified by the timescale of the Rayleigh-Taylor instability \citep[e.g.,][]{SOLOMATOV201581}{}{}. The instability leaves the  solid-melt magma mixture with a thermal structure that follows an adiabatic profile, and its timescale  is found to be ${\sim} 10^0 {-} 10^3 \ {\rm yr}$, that is, much shorter than the solidification timescale  \citep[e.g.,][]{miyazaki2019timescale,korenaga2023rapid}{}{}. We thus compute the temperature profile inside the convective mantle, when the temperature is above the liquidus or below the solidus, as an adiabatic and isentropic profile defined via \citep[e.g.,][]{turcotte2014geodynamics}{}{}:
\begin{equation}\label{mantle_adiabat}
    \frac{dT}{dP}\Bigg|_{T<T^{s}\,\vee\, T>T^{ l}}^{\rm ad}=\frac{\alpha T}{\rho C}
\end{equation}
Here, $P$ is the pressure, $T$ is the temperature, $\rho$ is the density, $\alpha$ is the thermal expansivity, and $C$ is the heat capacity, and we use the parameterized relations obtained in \citet{miyazaki2019timescale} and \citet[][]{korenaga2023rapid}{}{} for the material properties. In the two-phase mixture when mantle temperatures lie between the solidus and the liquidus, the adiabatic profile in \eq{mantle_adiabat} is modified to allow for the latent heat of fusion, thus we have:
\begin{equation}\label{mantle_adiabat2}
    \frac{dT}{dP}\Bigg|_{T^{ s} >T>T^{l}}=  \frac{dT}{dP}\Bigg|_{T<T^{s}\,\vee\, T>T^{l}}^{\rm ad}  + \frac{H_{\rm f}}{C}\gamma,  
\end{equation}
with $H_{\rm f}$ denoting the latent heat of fusion, linearly interpolated from $6\times10^5$~J~kg$^{-1}$ at the surface to
$9 \times10^6$~J~kg$^{-1}$ at the core-mantle boundary; and $\gamma$ denoting the extent of melting as a function of pressure \citep[][]{langmuir1992petrological}{}{}:
\begin{equation}
    \gamma = \left({\frac{dT}{dP}\Bigg|^{\rm ad}_{T<T^{ s}\,\vee\, T>T^{l}}-\frac{dT^{s}}{dP}}\right)\left({\frac{H_{\rm f}}{C} + \frac{dT}{dF_{\rm m}}}\right)^{-1}.
\end{equation}
The above equation thus controls the amount of melting produced upon pressure variation. The slope of the adiabat, the slope of the solidus, the  heat of fusion, the mantle heat capacity, and a proportionality function between temperature changes and produced melt are thus required to compute the pressure-dependence of melt. The outcome of solving the differential equations \eqref{mantle_adiabat} and \eqref{mantle_adiabat2} is shown in the phase diagram of Figure \ref{Fig_Mantle_adiabats}. The diagram is similar to those produced in \citet[][]{miyazaki2019timescale}{}{} and \citet[][]{korenaga2023rapid}{}{}. The key feature of this phase diagram is that, starting with a post-impact molten Earth that is cooling in time, a high temperature adiabat crosses the liquidus, and then the critical melt profile, in the lower mantle before it does in the upper mantle. Solidification thus starts at the bottom of the mantle leaving a surface magma ocean of shrinking thickness in time. This scenario of solidification leaves us with a setting of which tidal dynamics can be described with our formalism of Sections \ref{Section_Magma_isolated} and \ref{Section_magma_ocean_subsurface}. We acknowledge, however, that the analytical form of magma tides were developed in a thin layer approximation, which is violated for a fully molten mantle $(H>0.3 R_{\rm p})$; a limitation that we comment on further in the discussion section. 

\subsection{Early Earth tidal dissipation}\label{Section}
Dissipation in an early hot Earth, driven by the tidal forces of a close-in Moon is computed using \eq{Eq_dissipation}, via the dynamical Love number we provide in \eq{eq_full_LN}. The latter Love number self-consistently encodes for the tidal response a solid mantle, a magma ocean, and a solid surface crust. For the early Earth, the magma ocean extends to the surface, thus terms associated with crustal deformation (those including the membrane constant $\Lambda_n^{\rm M})$, are ignored.

For a specific surface temperature, we recover the thermal structure of the Earth's interior using the phase diagram of Figure \ref{Fig_Mantle_adiabats}, assuming a critical melt fraction $F_{\rm m, c}=0.4$. The thermal structure dictates the thickness of the solid silicate mantle, where $F_{\rm m}<F_{\rm m, c}$, and that of the magma ocean, where $F_{\rm m}>F_{\rm m, c}$. Tidal dissipation in the mantle is then computed using $\texttt{ALMA3}$ as described in Section \ref{Section_Tides_Solid}, assuming density and shear modulus profiles that follow the PREM model \citep[][]{dziewonski1981preliminary}{}{}, and a viscosity profile that follows the Arrhenius equation, modified to allow for the presence of partial melt in the solid matrix, namely:
\begin{equation}\label{viscosity_temperature}
    \eta(z,T) = \eta_{\rm s}(z)\exp\{-B F_{\rm m}\} \exp\left\{\frac{E_{\rm a}}{R T^{ s}(z)}\left(\frac{T^{ s}(z)}{T}-1\right) \right\},
\end{equation}
where $\eta_{\rm s}$ is the viscosity profile of the solid background, that is, at the solidus temperature, $B$ is an experimentally fitted parameter that ranges between 10 and 40 \citep[e.g., $B\sim 26$ for diffusion creep; ][]{mei2002influence}{}{}, $E_{\rm a}$ is the activation energy, and $R= 8.314462 \ {\rm J \ mol^{-1} \  K^{-1}}$ is the ideal gas constant \citep[][]{tiesinga2021codata}{}{}. We do not consider variations in the activation energy with depth, and we fix it to 
$300 \ {\rm  kJ \ mol^{-1}}$ \citep[][]{korenaga2006archean}{}{}. Once the material profiles are provided to $\texttt{ALMA3}$, the solid tidal and loading Love numbers are computed for any given tidal forcing frequency, assuming an Andrade rheology with $\alpha_{\rm A}=0.25$ and $\tau_{\rm A}=\tau_{\rm M}$ \citep[e.g.,][]{makarov2012conditions}{}{}. 

\begin{figure}[t!]
\includegraphics[width=.45\textwidth]{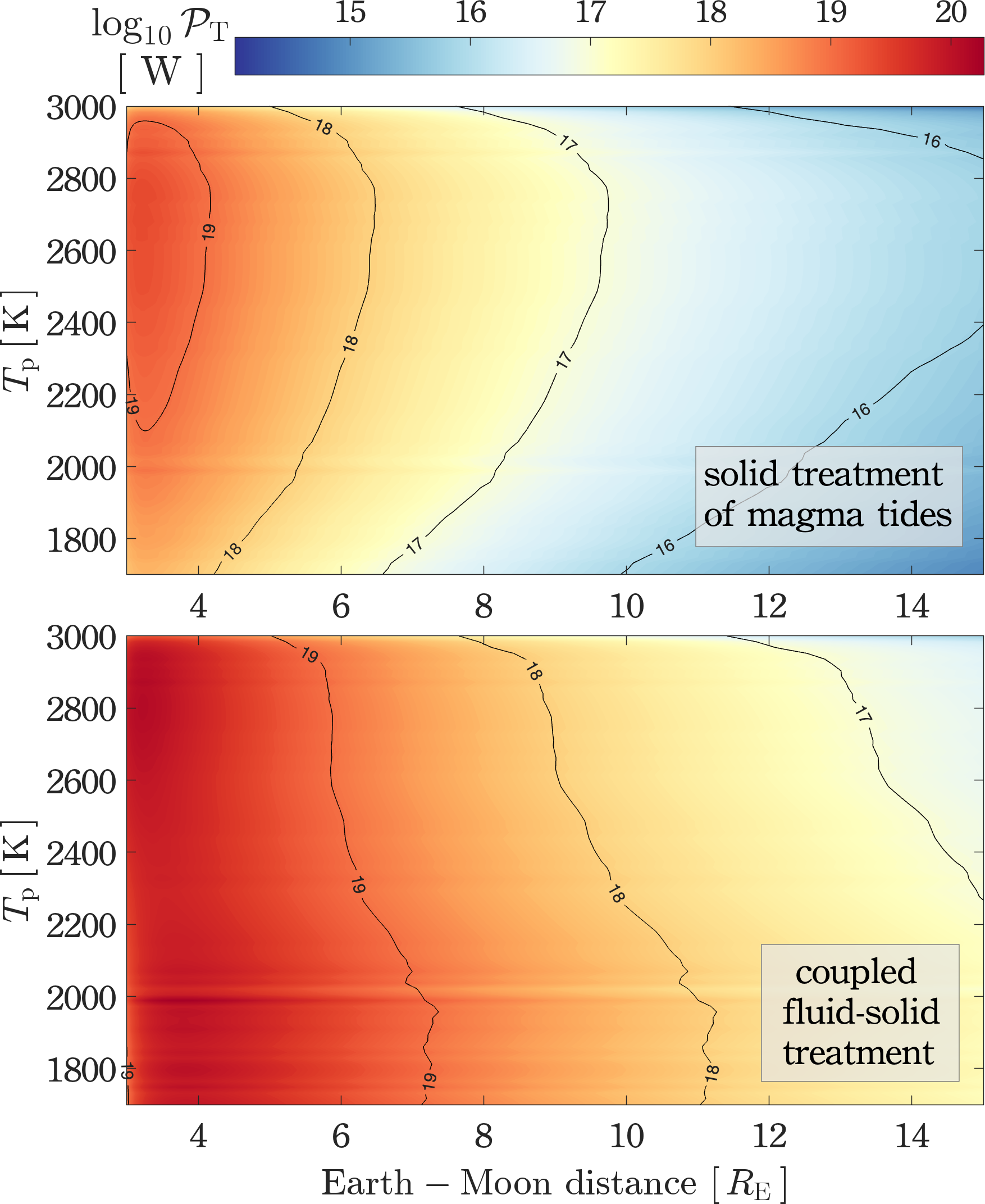}
\caption{Power dissipated by semi-diurnal terrestrial tides as a function of the mantle's potential temperature and the Earth-Moon distance. The top panel assumes that the magma ocean behaves like a low viscosity solid (i.e., at mantle depths where $F_{\rm m}>F_{\rm m,c},$ the viscosity takes the value of the melt viscosity $\eta_{\rm L}=0.1$ ${\rm Pa \ s }$). In the bottom panel, magma tides are recovered from the fluid treatment in Section \ref{Section_magma_ocean_subsurface}. The solid part of the mantle in both cases responds following an Andrade rheology.}
\label{Fig_Early_Earth_moon_diss}
\end{figure}

In the magma ocean layer, where the thermal structure lives above the profile of the critical melt fraction, we compute the tidal response using the analytical solution of the LTEs we provide in Section \ref{Section_Magma_isolated}. This is in contrast to all earlier studies where tidal dissipation in the Earth's magma layer is computed using the solid formalism, assuming a sharp drop in viscosity from a solid mantle  viscosity (${\sim} 10^{20} {-} 10^{23} \ {\rm Pa \ s}$) towards a melt viscosity (${\sim} 10^{-2} {-} 10^{1} \ {\rm Pa \ s}$) \citep[e.g.,][]{zahnle2015tethered,korenaga2023rapid}{}{}. 

In Figure \ref{Fig_Early_Earth_moon_diss},  we compute and plot the resulting tidal dissipation of the two approaches as a function of the Earth-Moon distance, and the Earth's potential temperature. In a circular coplanar setting, we focus on the semi-diurnal component of the tidal response. The top panel corresponds to the solid treatment of the magma layer, while the bottom shows our novel approach. The top panel is in good agreement with Figure 1 of \citet[][]{korenaga2023rapid}{}{}, although the solid tidal response was computed using two different codes and assuming different solid rheologies. The bottom panel shows that accounting for the fluid behavior of the magma ocean
significantly increases the dissipated power, by as much as one to two orders of magnitude, at any given temperature and tidal frequency. This confirms that the tidal response of the early Earth lives within the spectral region where fluid magma tides leave the notable tidal signature we discuss in Section \ref{Section_Magma_isolated}.

\subsection{Coupled thermal evolution of the Earth and Lunar orbital recession}\label{Section_coupled_EM}

After confirming that the fluid behavior of magma tides is relevant for the early Earth, we move now to investigate the effect of the magma ocean on the early evolution of the Earth-Moon system by coupling the thermal evolution of the Earth with the orbital evolution of the Moon. To isolate the effect of the magma ocean, and to better compare our results with earlier studies \citep[e.g.,][]{zahnle2015tethered,korenaga2023rapid}{}, we adopt a circular, coplanar orbital configuration for the Moon, which restricts tidal interactions in the system to the semi-diurnal component. We also focus on terrestrial tides, bearing in mind that Lunar tides at this early stage are significant and should be accounted for in a full dynamical model. Furthermore, for a very close-in Moon, Solar tides are negligible, leaving the total angular momentum of the Earth-Moon system conserved and 
\begin{equation}\label{orbital_evolution}
    \frac{d L_{\Omega}}{dt} = -\frac{dL_{\rm M}}{dt}  = -\mathcal{T}_{\rm M},
\end{equation}
where $\mathcal{T}_{\rm M}$ is the Lunar semi-diurnal tidal torque, $L_{\rm M}=\beta\sqrt{G(M_{\rm E}+M_{\rm M})a_{\rm M}}$  is the orbital angular momentum of the Moon, with $\beta=M_{\rm E}M_{\rm M}/(M_{\rm E}+M_{\rm M})$ denoting the system's reduced mass [$G(M_{\rm E}+M_{\rm M})=8.997\times 10^{-10}\ {\rm AU^3 \ day^{-2}}$; $M_{\rm E}/M_{\rm M}=81.3$; \citet{INPOP21a}], and $L_{\Omega} = C_{\rm E} (\Omega)\Omega$ is the rotational angular momentum of the Earth, with the velocity dependent principal moment of inertia given by \citep[e.g.,][]{goldreich1966history}{}{}:
\begin{equation}
    C_{\rm E}(\Omega) = C_{\rm E}(\Omega_0)  +\frac{2k_2^{\rm f}R_{\rm E}^5}{9G}(\Omega^2-\Omega_0^2).
\end{equation}
Here, $C_{\rm E}(\Omega_0)=0.3307M_{\rm E}R_{\rm E}^2$, and  $k_2^{\rm f}=0.93$ \citep[][]{spada2011benchmark,farhat2022resonant}{}{} is the Earth's fluid Love number of centrifugal deformation. The tidal torque, $\mathcal{T}_{\rm M}$, exerted about the spin of the Earth is computed in a similar way of computing the dissipated power of \eq{Eq_dissipation} \citep[see e.g.,][]{auclair2023can}{}{}:
\begin{equation}
\label{Eq_torque}
   \mathcal{T}_{\rm M} = \frac{R_{\rm p}}{8\pi G}\sum_{s=-\infty}^\infty\sum_{n=2}^\infty\sum_{m=0}^n m(2n+1)|U_{n,m}^{\rm T; s}|^2 \,\mathfrak{Im}\left\{k_{n}^m\right\}.
\end{equation}
Therefore, for the semi-diurnal tidal component, the torque is proportional to the dissipated power via $\mathcal{T}_{\rm M} = \mathcal{P}_{\rm T}/(\Omega-n_{\rm M})$, where $n_{\rm M}$ is the orbital mean motion of the Moon. 

In parallel, the thermal evolution of the Earth follows a balance between heat generated in its interior and that escaping through its surface. Internally generated heat includes the mantle's radiogenic heating, the flux generated by the core, and the tidally dissipated energy. Half-life times and  heat production rates of decaying radioactive isotopes allow us to safely neglect their contribution during the magma ocean's lifetime \citep[e.g.,][]{carlson2000timescales}{}{}.
Heat flow from the core depends on several variable parameters, the most important of which are the properties of the thermal boundary layer at the bottom of the mantle where heat is transferred from the core via conduction \citep[e.g.,][]{buffett2002estimates}{}{}. Estimates of this flow rate place it ${\sim} 10 \ {\rm TW}$  \citep[e.g.,][]{o2017thermal}{}{}, while end-member configurations of the boundary layer show that the thermal evolution, and consequently the lifetime, of the magma ocean is insensitive to the core heat flux \citep[][]{monteux2016cooling}{}{}. Therefore, we trace the thermal evolution of the Earth using a one-dimensional balance model between the tidally generated energy and the flux escaping the surface; namely \citep[e.g.,][]{christensen1985thermal}{}{}:
\begin{equation}\label{thermal_evolution}
    C_{\rm m}(T_{\rm p})\frac{dT_{\rm p}}{dt} =\mathcal{P}_{\rm T}(t)-Q_{\rm s}(t),
\end{equation}
where $ C_{\rm m}(T_{\rm p})$ is the mantle's heat capacity, $T_{\rm p}$ is the potential temperature, and $Q_{\rm s}$ is the geothermal heat flow at the surface due to the strongly convective magma ocean. We adopt the heat capacity function used in \citet[][]{korenaga2023rapid}{}{}, and we compute the convective heat flux via \citep[e.g.,][]{SOLOMATOV201581}{}{}:
\begin{equation}\label{convection_Soft_turbulence}
    Q_{\rm s} = 0.089 A_{\rm E} k_{\rm L} \frac{\left(T_{\rm p}-T_{\rm s}\right)}{H} Ra^{1/3}.
\end{equation}
Here, $A_{\rm E}$ is the surface area of the Earth, $k_{\rm L} = 2~{\rm W~K^{-1}~m^{-1}}$. is the thermal conductivity of the melt, $T_{\rm s}$ is the surface temperature, parameterized, following  \citet[][]{miyazaki2022wet}{}{} and \citet[][]{korenaga2023rapid}{}{}, as $T_{\rm s}=546+0.63 T_{\rm p}$, and $Ra$ is the Rayleigh number, defined as:
\begin{equation}\label{Rayleigh_number}
    Ra = \frac{\alpha_{\rm s}g\left(T_{\rm p}-T_{\rm s}\right) H^3}{\kappa \eta_{\rm L}}
\end{equation}
where $\alpha_{\rm s}=5 \times 10^{-5}~{\rm K^{-1}}$ is the thermal expansivity, $\kappa = k_{\rm L}/(\rho_{\rm L}C_{\rm p})$ is the coefficient of thermal diffusivity, with $\rho_{\rm L} = 3000{\rm~kg~m^{-3}}$ being the melt's density, and $C_{\rm p}=103{\rm ~J~kg^{-1}~K^{-1}}$ denoting the specific heat \citep[][]{korenaga2023rapid}{}{}. {Strictly under these assumptions, the major variable in the convective flux is thus the melt's viscosity, $\eta_{\rm L}$ (we comment further on the adopted temperature parameterization in the discussion section).} Experimental measurements and theoretical constraints estimate the viscosity of near-liquidus peridotitic melt at low pressures to be ${\sim} 0.01 {-} 0.1~{\rm Pa~s}$ \citep[e.g.,][]{dingwell2004viscosity,liebske2005viscosity}{}{}. But these values increase with lower temperatures and  lower degrees of partial melting to 100 Pa s \citep[e.g.,][]{kushiro1986viscosity,rubie2003mechanisms}{}{}. We thus treat $\eta_{\rm L}$ as a free parameter encompassing this range. 
\begin{figure}[t!]
\includegraphics[width=.45\textwidth]{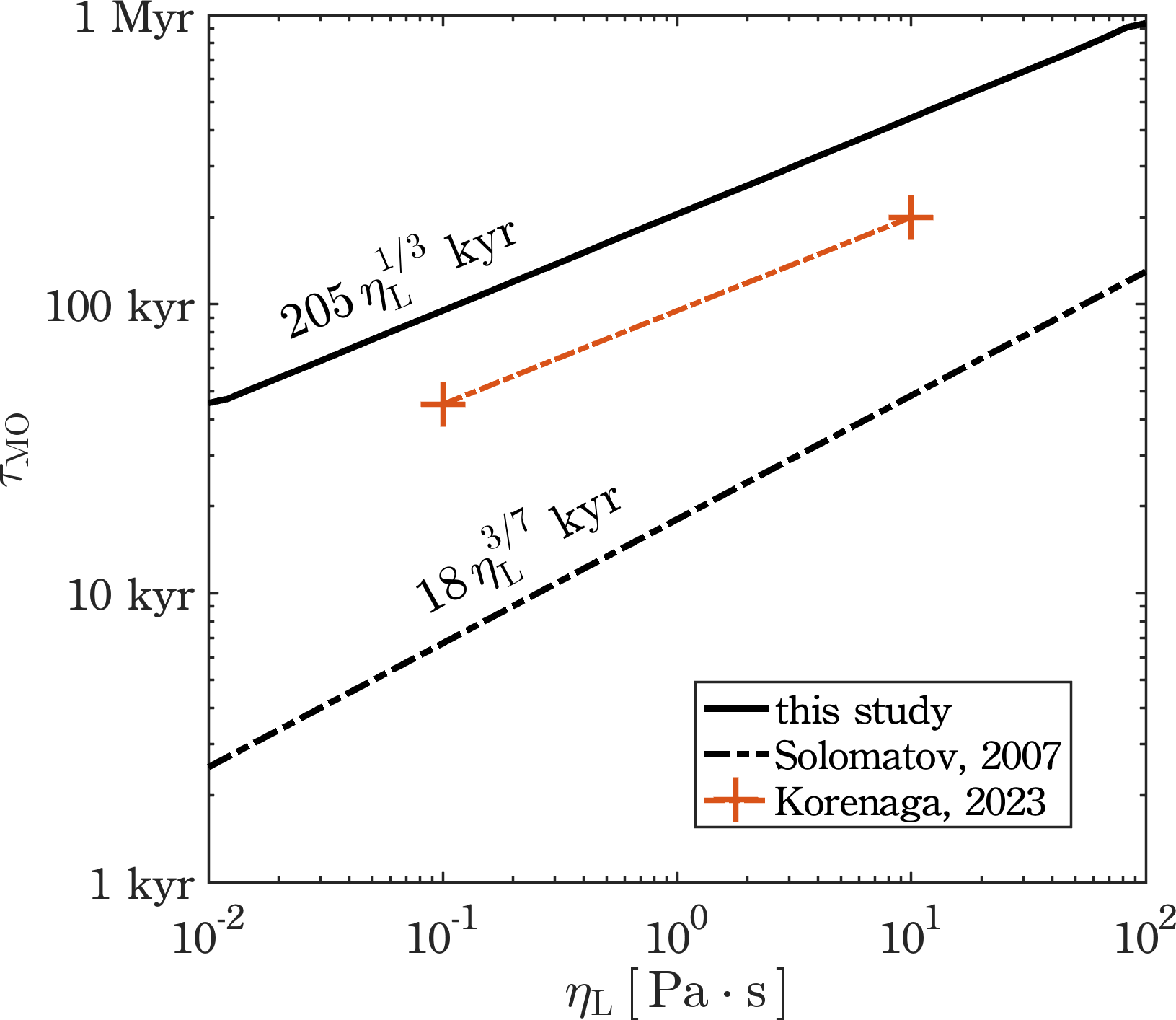}
\caption{The lifetime of the Earth's {fluid} magma ocean, $\tau_{\rm MO}$, as a function of the melt viscosity, $\eta_{\rm L}$. {We define the termination of the fluid magma phase by the time at which the surface temperature drops below 1610~K.} The results of our simulations, where we assume soft turbulence and we account for the fluid behavior of magma tides is shown by the solid line. The dashed line corresponds to the results of \citet[][]{solomatov2007magma}{}{}, where hard-turbulence convection is modeled, but tidal heating is ignored. The red line interpolates the two points obtained in \citet[][]{korenaga2023rapid}{}{} at $\eta_{\rm L}=0.1$, 100~{Pa~s}.}
\label{Fig_MO_lifetime}
\end{figure}

We simulate the early evolution of the Earth-Moon system by solving the coupled system of orbital-thermal evolution (\ref{orbital_evolution}-\ref{thermal_evolution}). We initiate the system with a Moon having its present mass, right outside the Roche limit ($a_{\rm M;0}=3.5 R_{\rm E})$. Lunar accretion occurs over relatively very fast timescales \citep[${\sim}1{-}100 \ {\rm yrs}$, e.g., ][]{thompson1988gravitational,salmon2012Lunar}{}{}, but interactions between the accreting moon and the circumplanetary debris disc can push the Moon towards $\sim 5R_{\rm E}.$  We also start the evolution with a terrestrial potential temperature of 4500~K. We have confirmed, however, that the results we present here are insensitive to variations in these initial conditions, as long as we assume angular momentum conservation and that the initial temperature is well above the liquidus.

As the Earth starts cooling, the magma ocean starts solidifying from the lower mantle. This is due to the stronger slope of the liquidus compared to that of the mantle adiabat in the phase diagram of Figure \ref{Fig_Mantle_adiabats}. The magma ocean ceases to exist, in our simulations, when the mantle potential temperature reaches the surface value of the critical melt temperature profile, that is $T^{40}(z=0)$, which corresponds to ${\sim}1610~{\rm K}$. The residual melt in the subsurface magma slush would then start migrating radially upwards and degas, allowing for the formation of the Earth's water oceans. As we are interested here in isolating the effect of fluid-driven tidal heating in the Earth's magma ocean, we terminate our simulations right upon solidification. 


In Figure \ref{Fig_MO_lifetime}, we show the dependence of the {fluid} magma ocean's lifetime, $\tau_{\rm MO}$, on the melt viscosity, $\eta_{\rm L}$. Our simulations suggest a relation of the form: $\tau_{\rm MO}=205 \eta_{\rm L}^{1/3}~{\rm kyr}$. As such, for a near-liquidius, ultramafic silicate magma ocean, solidification occurs over a timescale on the order of $50{-}60 \ {\rm kyrs}$. If the ocean starts with lower degrees of melting, its lifetime is on the order of $400{-}500 \ {\rm kyrs}$. While this relation between the degree of melting and the magma ocean's lifetime could appear as counter-intuitive, it is essentially by virtue of lower temperatures driving weaker convection, thus attenuating the flux escaping at the surface, and delaying the ocean's complete solidification.

In the same Figure \ref{Fig_MO_lifetime}, we also plot the dependence on the melt viscosity obtained in \citet[][]{solomatov2007magma}{}{} and \citet[][]{korenaga2023rapid}{}{}. The former indicates a relation of the form: $\tau_{\rm MO}=18 \eta_{\rm L}^{3/7} $ kyr. This relation was also obtained by \citet[][]{monteux2016cooling}{}{} via direct numerical simulations. The difference between this dependence and ours is, to first order, due to the fact that we have modeled convection in the ordinary -- or as often called, the soft -- turbulence regime. Namely, for very high Rayleigh numbers ($Ra\geq 10^{19}$, Eq. \ref{Rayleigh_number}), convection is expected to enter a regime of hard turbulence, whereby the convective flux becomes stronger than that described by Eq. \ref{convection_Soft_turbulence} \citep[e.g.,][]{siggia1994high}{}{}. For a magma ocean occupying the larger volume of the mantle, the Rayleigh number can indeed be associated with hard turbulence. Describing this regime accurately, however, requires additional experimental work that is yet to be performed, thus regime extrapolation and scaling laws have been proposed \citep[e.g.,][]{castaing1989scaling,shraiman1990heat,roche2020ultimate}{}{}. 

On the other hand, the difference between our $\tau_{\rm MO}(\eta_{\rm L})$ and that we interpolate from the results of \citet[][]{korenaga2023rapid}{}{} can be primarily attributed to the effect heating driven by fluid tides. The almost two-order-of-magnitude amplification of tidal dissipation obtained in Figure \ref{Fig_Early_Earth_moon_diss},  due to fluid tides, increases $\tau_{\rm MO}$ by a factor of 2. This said, having modeled convection in the soft turbulence regime, and allowed for the additional tidal heating attributed to the fluid behavior of the magma ocean, our results can be considered as an upper limit of the {fluid} magma ocean's lifetime, the maximum of which, corresponding to a melt viscosity of 100 Pa s, is $\sim1$ Myr.  

\begin{figure}[]
\includegraphics[width=.45\textwidth]{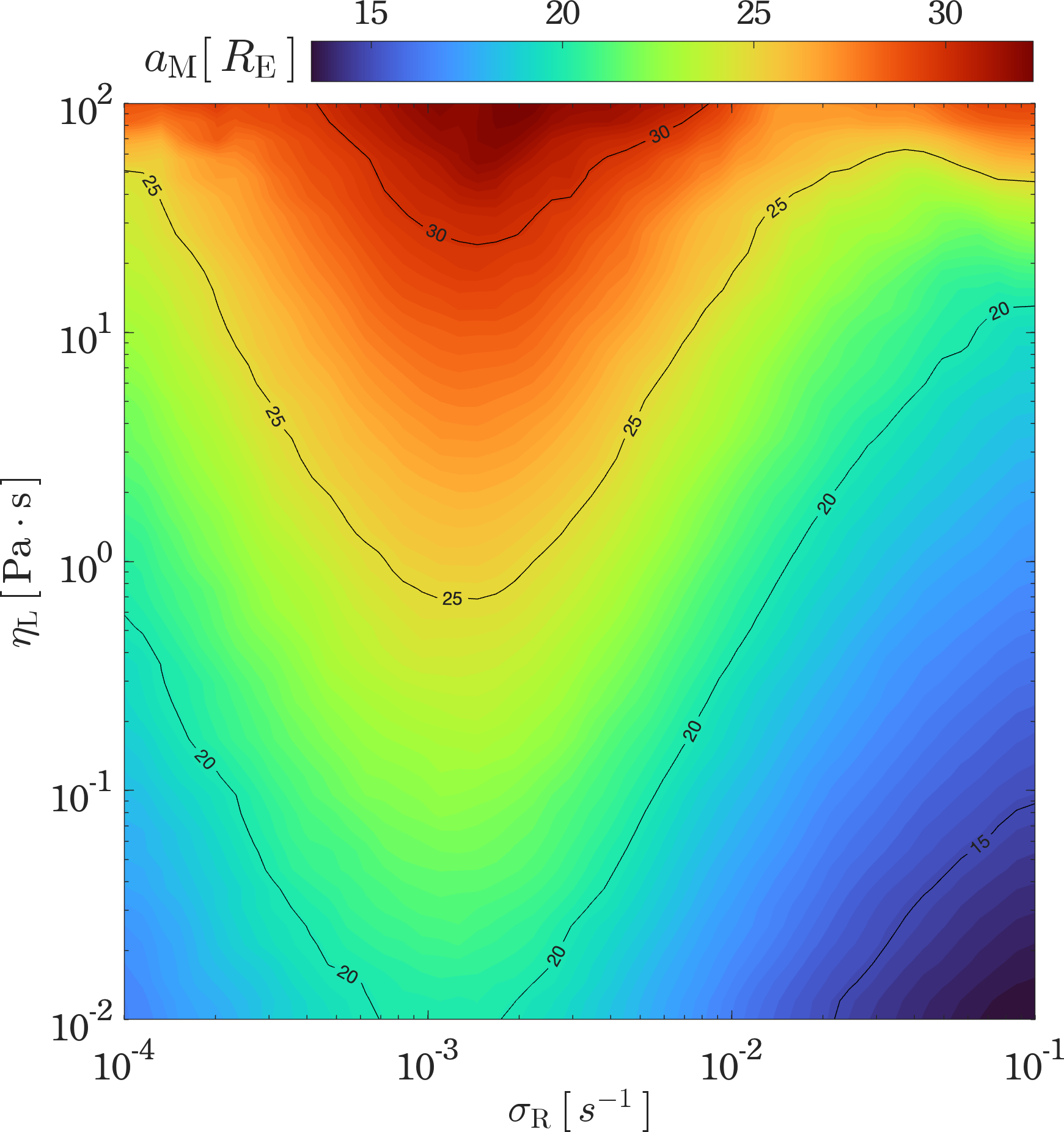}
\caption{Earth-Moon distance at the time of the Earth's magma ocean complete solidification. {The atmosphere throughout the evolution is assumed to be relatively thin and dry (see the main text)}. The contour surface marks the terminal state of our thermal-orbital coupled simulations of the Earth-Moon system described in Section \ref{Section_coupled_EM}.  The Lunar semi-major axis, $a_{\rm M}$, is recovered from the simulations for ranges of the magma ocean dissipative frequency, $\sigma_{\rm R}$, and the mantle's melt viscosity $\eta_{\rm L}$. }
\label{Fig_Earth_Moon_Distance}
\end{figure}

Since the cooling of the magma ocean is primarily driven by the convective source {(we emphasize that we have assumed, following \citet{korenaga2023rapid}, a dry atmosphere for which thermal blanketing is negligible; the effect of the latter can be seen in the results of \citet{zahnle2015tethered})}, we confirm that the effect of tidal heating on the Earth's magma ocean lifetime is secondary, even when allowing for the fluid tidal behavior. However, this does not extrapolate to the orbital evolution of the Moon, where terrestrial tides are the main dynamical player. In Figure \ref{Fig_Earth_Moon_Distance}, we plot the Earth-Moon distance, $a_{\rm M}$, reached upon the magma ocean solidification stage, as a function of the melt viscosity and the oceanic dissipative frequency $\sigma_{\rm R}$. We consider a range of $\sigma_{\rm R}$ that guarantees the creep flow tidal regime adopted to describe the tidal response of the magma ocean (Section \ref{Section_Magma_isolated}). Moving vertically upwards in the plot and increasing $\eta_{\rm L}$, we extend the lifetime of the magma ocean, and consequently the time window within which the Moon is subject to the amplified tidal response of the fluid. This effect accelerates Lunar recession, driving the Moon towards a high orbit of $\sim 30 R_{\rm E}$ within less than 1 Myrs. In contrast, the variation of $a_{\rm M}$ as a function of $\sigma_{\rm R}$ is non-monotonic. This is by virtue of the dependence of the spectral position of the fluid tidal peak on $\sigma_{\rm R}$, as seen in Figure \ref{Fig_love_number_magma}. Namely, it happens that for $\sigma_{\rm R}\approx 10^{-3}$ s$^{-1}$, along with the corresponding dynamical cooling history of the Earth, the peak of the fluid tidal response occurs exactly in the spectral region of the early Earth-Moon dynamics. This explains the higher Lunar orbit reached around this value of $\sigma_{\rm R}$. 

Our parametric study further indicates that the fluid response of the magma ocean guarantees a minimum $a_{\rm M}$ of 14$R_{\rm E}$ upon the complete solidification of the Earth. This, however, corresponds to a narrow region of the parameter space; that is, it is more likely that the Moon must have reached $20 {-} 25~R_{\rm E}$ within the first million years of its lifetime. Our estimates thus fall in between the limited Lunar recession to $a_{\rm M}\sim7-9 R_{\rm E}$ obtained by \citet[][]{korenaga2023rapid}{}{}, due to their solid description of magma tides, and the boosted Lunar recession to $a_{\rm M}\sim 40 R_{\rm E}$ obtained by \citet[][]{zahnle2015tethered}{}{}, due to {the delayed solidification of the Earth driven by atmospheric thermal blanketing}. More stringent constraints on the two used free parameters are therefore much needed to better predict the state of the Lunar orbit right upon terrestrial solidification.

\section{Tidal Dissipation for Close-in Rocky Exoplanets}\label{Section_Close_In_Exo}
Moving afar from home, we explore in this section the effect of fluid tides in a magma layer on the thermal and spin-orbit dynamics of close-in exoplanets,  starting with fiducial planets, before switching to the observed population. {The interplay between tidal heating and the spin-orbit state of a planet has a profound influence on its climatic regime, and consequently its prospects for habitability \citep[e.g.,][]{dobrovolskis2007spin,dobrovolskis2009insolation,barnes2013tidal}.} For the purpose of characterizing this effect uniquely, we consider an isolated star-rocky planet system. Planets living well within the orbit of Mercury, specifically on an orbital period $\leq 20$~days  are amenable to sustain a molten surface considering the extreme thermal forcing they receive from their host stars and the internal heat generated by the stellar tidal forcing. Planets further out on Earth-like orbits can also harbor a magma ocean as a byproduct of any impact scenario. 
\subsection{The effect of magma oceans on planetary spin-orbit dynamics}
In our fiducial star-planet system, we explore the effect of fluid tides in the putative magma ocean on the secular (long-term) spin-orbit dynamics of the planet. We focus on terrestrial tides, ignoring  tides within stellar fluid envelopes that may in some regimes control the tidal evolution of the system, or any magnetic interactions \citep[e.g.,][]{bolmont2012effect,ogilvie2013tides,ferraz2023tidal}{}{}. We presume that a planet starts in a thermal equilibrium state with a temperature profile that allows for the existence of a magma ocean. As the planetary orbital and rotational variables tidally evolve, the thermal equilibrium would also shift, but we presume that the planet adiabatically follows the high temperature stable equilibrium, thus maintaining its magma ocean.

In the absence of any other tidal players (such as atmospheric tides), and if the planet is to avoid merger with the star, the planet tidally evolves towards an equilibrium state of spin-orbit synchronization on a circular orbit, with its spin and orbital angular momentum vectors aligned \citep[e.g.,][]{mignard1980evolution,hut1981tidal}{}{}. The typical timescales associated with these evolutions, however, are highly dependent on the planetary rheology, with earlier studies mapping out this dependence \citep[e.g.,][]{jackson2008tidal,matsumura2010tidal,correia2014deformation,boue2016complete,renaud2018increased}{}{}. The timescales in question are of significance for the survival of close-in  exoplanets since the faster they undergo orbital circularization the safer they are from close encounters, scattering, and dynamical resonances that may lead to ejection or destruction \citep[e.g.,][]{matsumura2013effects,mustill2015destruction,touma2015disruption,pu2018eccentricities}{}{}. One then naturally wonders what is the effect of fluid tidal heating within a magma ocean on the typical tidal evolution and the timescale of circularization of a close-in exoplanet. 

Considering a planar configuration, that is, ignoring planetary and stellar obliquities, the planet's orbital semi-major axis, $a$, eccentricity, $e$, and rotational velocity, $\Omega$, evolve secularly under terrestrial tides following \citep[][]{correia2022tidal}{}{}:
\begin{align}\label{Omega_evo}
    \dot{\Omega} = \frac{\mathcal{I}_0}{\mathcal{C}}\!\!\sum_{s=-\infty}^{\infty} -\frac{3}{2}\mathcal{K}_{2,s}\left(X_s^{-3,2}(e)\right)^2. 
\end{align}
\begin{align}\nonumber
    \dot{e} &= \mathcal{E}_0 \frac{\sqrt{1-e^2}}{4e} \!\!\sum_{s=-\infty}^{\infty}\Bigg\{ \!\!- \mathcal{K}_{0,s}\left(X_s^{-3,0}(e)\right)^2 \!s\sqrt{1-e^2} \\\label{eccentricity_evo}
&+3\,\mathcal{K}_{2,s}\left(X_s^{-3,2}(e)\right)^2\left(2-s\sqrt{1-e^2}\right) \Bigg\}, 
\end{align}
\begin{align}
    \frac{\dot{a}}{a} = \mathcal{E}_0\!\!\sum_{s=-\infty}^{\infty} \frac{s}{2} \left[- \mathcal{K}_{0,s}\left(X_s^{-3,0}(e)\right)^2+3\,\mathcal{K}_{2,s}\left(X_s^{-3,2}(e)\right)^2 \right],
\end{align}
In these equations, $\mathcal{K}_{p,q} =\mathfrak{Im} \{ k_2(p\Omega - qn_{\rm orb}\}$, 
$\mathcal{E}_0 = n_{\rm orb}\left(\frac{M^\prime}{M_{\rm p}}\right)\left(\frac{R_{\rm p}}{a}\right)^5$,  $\mathcal{I}_0=GM^\prime R_{\rm p}^5/a^6 $, and $\mathcal{C}$ is the principal moment of inertia of the planet. The timescales of the evolution of these three elements are well separated, and they can be ranked from shortest to longest following the order of the equations. This poses a computational difficulty if one were to explore a wide parameter space by direct numerical integrations. Therefore, here we recover instead the typical tidal dynamics of the planet and the effect of the magma ocean thereof by mapping the structure of tidal equilibria \citep[e.g.,][]{ragazzo2024tidal}{}{}. Namely, we consider the tidal equilibria living on the torque-free curve defined by freezing the rotational velocity in \eq{Omega_evo}:
\begin{equation}\label{Eq_torque_free}
\mathcal{T}_{f}\coloneq \left\{\left(e,\Omega,n_{\rm orb}\right): \sum_{s=-\infty}^{\infty}\mathcal{K}_{2,s}\left(X_s^{-3,2}(e)\right)^2=0\right\}.
\end{equation}

\begin{figure}[t]
\includegraphics[width=.47\textwidth]{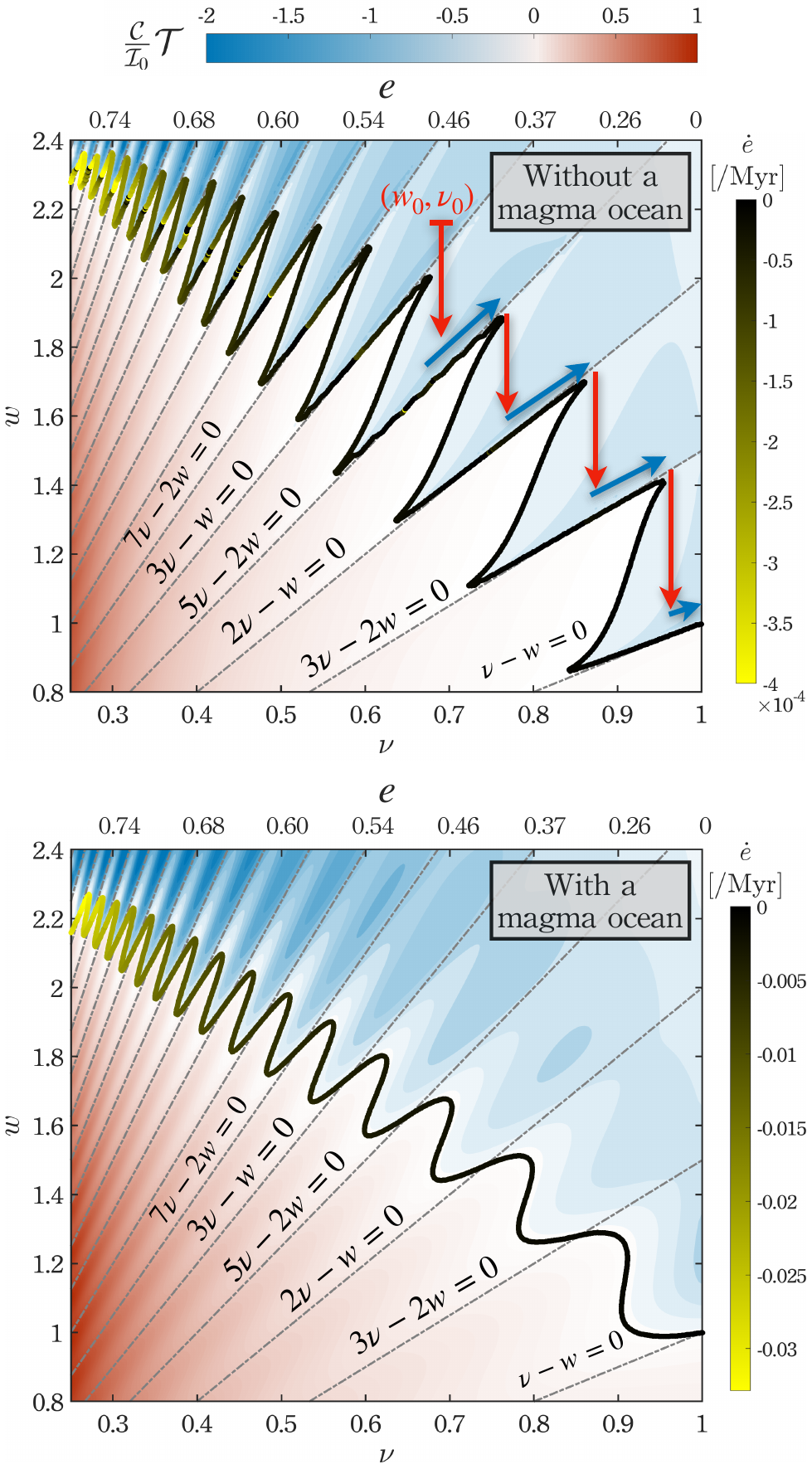}
\caption{Spin-orbit dynamics of a tidally forced planet, with (top panel) and without (bottom panel) a magma ocean. The nondimensional coordinates are $\nu=(1-e^2)^{3/2}$ and $w =(1-e^2)^{3/2} \Omega/n_{\rm orb}$, thus moving to the right corresponds to energy dissipation and orbital circularization and  $(w,\nu)=(1,1)$ marks spin-orbit synchronization. Plotted is a contour surface of the tidal torque (nondimensionalized) governing the rotational evolution of the planet (Eq. \ref{Omega_evo}), with the blue region marking a negative torque (decelerative effect), and the red region marking a positive torque (accelerative effect). The highlighted curve corresponds to tidal relative equilibria where $\dot{\Omega}=0$ (Eq. \ref{Omega_evo}) as defined by the torque free curve of \eq{Eq_torque_free}. The curve is color coded with the rate of eccentricity decay of these equilibria, computed via \eq{eccentricity_evo}. We note the difference of scales between the colorbars of the two panels. Dashed lines mark spin-orbit resonances.}
\label{Fig_Torque_Free}
\end{figure}

A contour surface of the tidal torque is plotted in Figure \ref{Fig_Torque_Free} for two cases: with and without a magma ocean. The typical spin-orbit dynamics are best portrayed via the chosen non-dimensional coordinates: $\nu=(1-e^2)^{3/2}$ and $w =(1-e^2)^{3/2} \Omega/n_{\rm orb}$. As such, $(w,\nu)=(1,1)$ corresponds to the equilibrium state of spin orbit synchronization, while higher order spin-orbit resonances are highlighted with the dashed lines and labeled accordingly (for instance, the $2\nu-w=0$ corresponds to the 2:1 spin-orbit resonance with varying eccentricity). The blue (red) part of the surface marks the parametric region where the torque is negative (positive), depleting (increasing) the rotational angular momentum of the planet. 

Consider a magma-free planet starts its dynamical endeavour with $(w_0,\nu_0)$ as indicated in the top panel. The negative torque generated by planetary solid tides rapidly despins the planet towards a tidal equilibrium on the torque-free curve (in black) with the closest eccentricity value, as indicated with the red downwards arrow. The planet is then captured into a quasi-equilibrium corresponding to the  5:2 spin-orbit resonance. Thereafter,  it evolves adiabatically, and very slowly, on the torque free curve while decreasing its eccentricity (increasing $\nu$) following the blue arrow. It continues to do so until a fold point in the equilibrium structure is reached, whereby tidal energy dissipation takes over and forces the planet to escape the resonance lock, resulting in a relatively abrupt spin jump toward a lower order spin-orbit resonance. This cascade of surfing resonances continues with the associated fast and slow evolution timescales until the planet reaches its destined spin-orbit synchronization. 

Adding a magma ocean to the planet distorts this standard picture. As we show in the lower panel of Figure \ref{Fig_Torque_Free}, the torque-free curve is no longer -- intermittently -- aligned with the dashed lines, indicating that the tidal quasi-equilibria do not correspond to the spin-orbit resonances anymore. A planet starting with the same initial conditions as in the top panel would still despin towards the closest equilibrium, but then it will cross the spin-orbit resonance without getting captured in it. This is by virtue of the strong fluid tides within the magma layer, which render the depletion of orbital energy much more efficient. As such, the orbital semi-major axis and eccentricity decay much faster than in the absence of the magma layer, preventing the resonance capture. Consequently, the time window within which the planet is subject to the slowly evolving resonant dynamics is diminished, accelerating the planet's track towards the final equilibrium of spin-orbit synchronization and orbital circularization. 

\begin{figure}[]
\includegraphics[width=.45\textwidth]{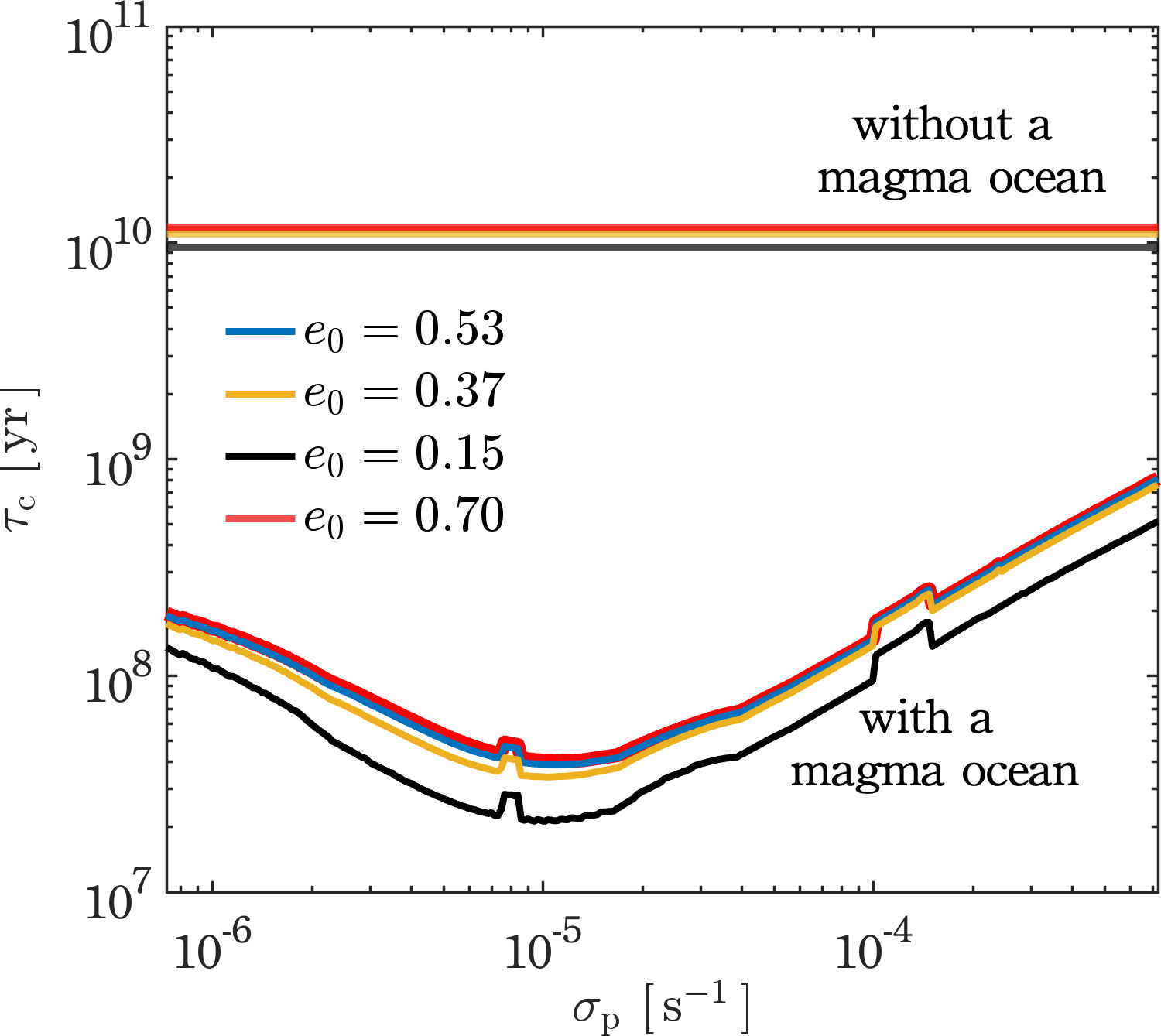}
\caption{Orbital circularization timescale, $\tau_{\rm c}$ as a function of the frequency of peak fluid{-driven} dissipation, $\sigma_{\rm p}$ (Eq. \ref{peak_frequency}). We only simulate planetary tides, ignoring the effect of tides raised within the star, of an Earth-like planet orbiting a $1 M_{\odot}$ star at 0.1 AU, with and without a magma ocean, and for different initial eccentricities. The frequency $\sigma_{\rm p}$ maps possible variations in the magma ocean thickness and dissipation timescale, or the planetary radius and mass.}
\label{Fig_tau_c}
\end{figure}

Additionally, the enhanced fluid tidal response directly affects the evolution of eccentricity through the Love number (Eq. \ref{eccentricity_evo}), further accelerating the rate of circularization. In Figure \ref{Fig_Torque_Free},
we color code the torque-free curve, mapping the rate of change of eccentricity for each tidal equilibrium. Evidently, the presence of a magma ocean forces the orbital eccentricity to decay at much faster rates than in its absence. However, regardless of the magma ocean's existence, the rate of change of eccentricity is higher for more eccentric orbits. This is due to the fact that the larger the eccentricity is the larger the contribution of higher order Hansen coefficients is in \eq{eccentricity_evo}. Computing these rates, one can then directly estimate the timescale of circularization, $\tau_{\rm c}$, of an initially eccentric orbit, by tracing the typical evolution we described in both cases. An example that contrasts the two cases is  plotted in Figure \ref{Fig_tau_c}. Therein, we compute $\tau_{\rm c}$ for an Earth-like planet at 0.1 AU from a $1 M_{\odot}$ star, as a function of $\sigma_{\rm p}$, the spectral position of fluid tidal response peak (Eq. \ref{peak_amp}). The latter mimics variations in both the magma ocean thickness, $H$, and the dissipative timescale, $\sigma_{\rm R}$. As such, $\tau_{\rm c}$, in the absence of a magma ocean, is not affected by these variations, and is held fixed around ${\sim} 10~{\rm Gyrs}$ with a weak  dependence on the initial eccentricity, $e_0$. In contrast, adding a magma ocean reduces $\tau_{\rm c}$ by 1-3 orders of magnitude, depending on $\sigma_{\rm p}$, or equivalently, on the spectral overlap between the tidal frequency of the system and $\sigma_{\rm p}$. We emphasize here that these timescales can get shorter by an additional 1-2 orders of magnitude in specific regimes upon the inclusion of stellar tides \citep[e.g.,][]{jackson2008tidal}{}{}. An Earth-like magma ocean-hosting exoplanet at 0.1 AU can thus completely circularize its orbit within few million years. 

\subsection{Implications for detected close-in, rocky exoplanets}
\subsubsection{Population compilation}\label{Section_population_compilation}
So where does this analysis leave us when considering the already detected close-in exoplanets? To illustrate the possible effect of  magma oceans on such planets, we compile from \url{http://exoplanet.eu} a list\footnote{As of April 2024.} of close-in Earth-like and super-Earth planets ($M_{\rm p}\leq10 M_{\oplus}, R_{\rm p}\leq 2R_{\oplus})$ orbiting with $P_{\rm orb}\leq 20$ days. The list comprises 108 planets, 59 of which have non-zero eccentricity, and thus constitute our sample at focus as they are subject to tidal heating from eccentricity tides (unless stated otherwise, we assume that these planets are spin-orbit synchronized, and thus the absence of semi-diurnal tides). {The parameters of the compiled population are displayed in Table \ref{Table_Exo}.} Out of these 59, 40 planets have $R_{\rm p}\leq 1.6 R_{\oplus}$ and are thus most likely consistent with a mostly rocky composition \citep[e.g.,][]{valencia2006internal,weiss2014mass,dorn2015can}. 

It is noteworthy that the average eccentricity of these planets is $\langle e\rangle=0.0912$. {Taking the eccentricity values at face value, if these planets indeed harbor shallow magma oceans (more on that in the next section), the latter observation naturally appears to be conflicting with the shorter orbital circularization timescales that  we report on in the previous section, therefore suggesting that we are overestimating the rate of fluid tidal dissipation. While this might be possible, in the absence of constrained ages for all of these planets, it remains elusive to specify whether the origin of these eccentricities is the systems' young age compared to their circularization timescales, or the effect of other dynamical interactions. The latter can include, for instance: \textit{i)} the possible history of vigorous scattering \citep[e.g.,][]{ford2008origins,matsumura2013effects,petrovich2014scattering}{}{}; \textit{ii)} secular excitations from giant planets and stellar companions residing on wide inclined and eccentric orbits \citep[e.g.,][]{mardling2007long,pu2018eccentricities,farhat2021laplace}{}{}; \textit{iii)} angular momentum exchange between the planet and a debris disk surrounding it \citep[e.g.,][]{terquem2010eccentricity,poblete2023self}; \textit{iv)} and possibly, a tidally-induced pumping of eccentricity in the presence of other gravitational interactions or stellar spin-orbit misalignment \citep[e.g.,][]{correia2011pumping,albrecht2012obliquities}. It is also possible that some of these close-in planets are in a fully fluid equilibrium state, such that dissipation is reduced and the circularization timescale is longer than predicted earlier. It is thus necessary to sufficiently characterize the dynamical architecture of the systems hosting each of these planets in order to explain their individual eccentricities.}

For each of the planets in our list, we estimate the total energy budget accounting for the effect of tidal heating, $\mathcal{P}_{\rm T}$, insolation, $\mathcal{P}_{\rm ins}$, and radiogenic heating, $\mathcal{P}_{\rm RG}$. Namely:
\begin{equation}
    \mathcal{P}_{\rm total} = \mathcal{P}_{\rm T}  + \mathcal{P}_{\rm ins}+\mathcal{P}_{\rm RG}.
\end{equation}
Tidal heating follows from \eq{Eq_dissipation}, taking into account the observed orbital parameters of each planet, and assuming an Andrade rheology for the solid interior with an Earth-like interior physical structure. 

The averaged insolation heating impinging on the surface, $\mathcal{P}_{\rm ins}$, is computed via
\begin{equation}\label{insolation_heating}
    \mathcal{P}_{\rm ins} = \pi R_{\rm p}^2 (1-A) \frac{L_{\star}}{4\pi a^2}.
\end{equation}
{In this equation, $A$ is the planetary  albedo \citep[$A\sim0.1$ for molten surfaces; ][]{essack2020low}, $L_{\star }$ is the stellar luminosity, and $a$ is the planet's semi-major axis. The stellar luminosity is computed, when possible, via the Stefan-Boltzmann law: $L_\star =4\pi R_\star^2\sigma_{\rm SB} T_\star^4$, where $R_\star$ and $T_\star$ are the stellar radius and temperature, respectively, 
and $\sigma_{\rm SB}= 5.6704\times 10^{-8}$~W m$^{-2}$ K$^{-4}$ is the Stefan-Boltzmann constant. 

As for the radiogenic component, we estimate it for each planet by scaling the heat production rates per kg of mantle mass, provided by \citet[][]{frank2014radiogenic}{}{}, due to the radioactive activity of the isotopes $^{232}$Th, $^{40}$K, $^{235}$U, and $^{238}$U. We assume that the mantle volume of each planet is proportionally similar to that of the Earth \citep[$\sim 84\%$ of the planetary volume, e.g., ][]{stacey2008physics}{}{}, and that the planet's age is that of its host star. However, for all the close-in planets in our compiled list with known stellar ages, we obtain $\mathcal{P}_{\rm ins}/\mathcal{P}_{\rm RG}\in[10^4,10^9]$ , and $\mathcal{P}_{\rm \rm T}/\mathcal{P}_{\rm RG}\in[10^2,10^8]$. It is thus safe to ignore the heating contribution of radionuclides in what follows. 

\begin{figure*}[ht]
\includegraphics[width=\textwidth]{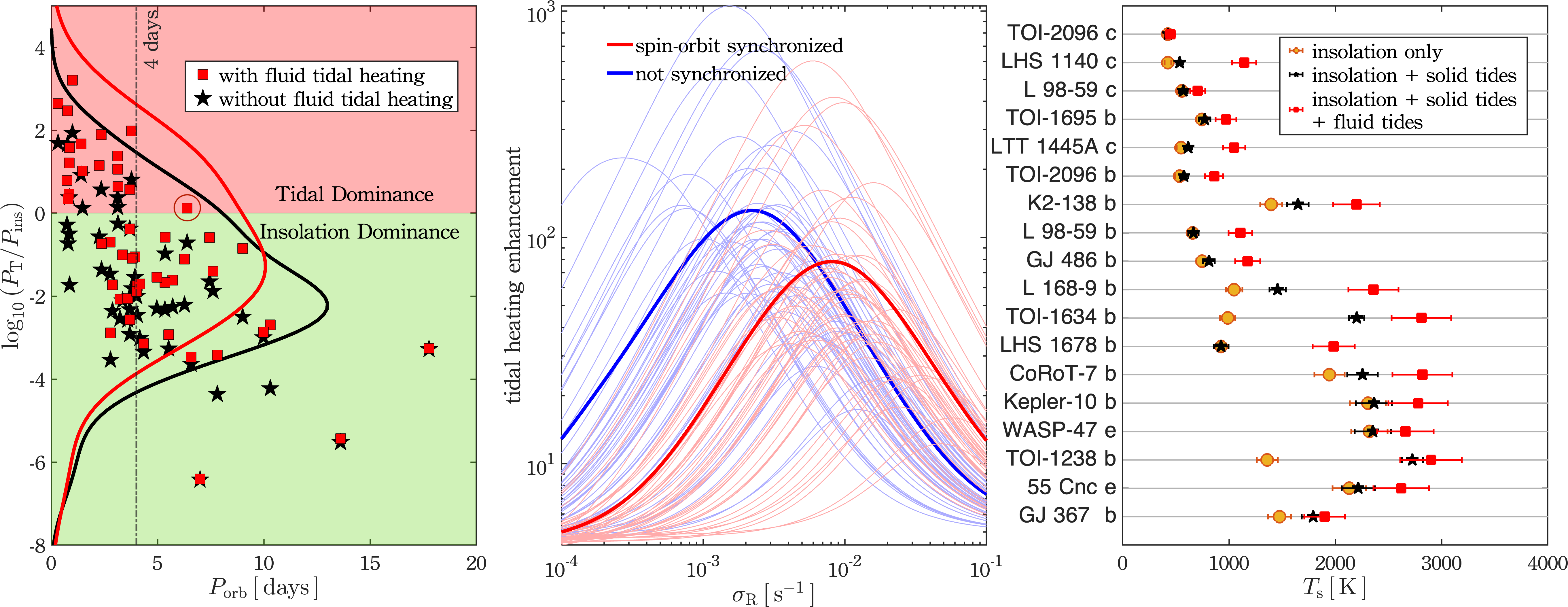}
\caption{Modeled heating and estimated surface temperatures of close-in exoplanets. \textit{Left:} Planetary heating via insolation is compared to tidal heating for a population of 59 close-in rocky exoplanets with non-zero eccentricity (see the main text for a description of the population). Tidal heating is computed in two cases: with (red squares) and without (black stars) fluid tides within a surface magma ocean on each planet. Solid curves depict the population distribution in each case. Noticeable is the threshold of $P_{\rm orb}=4~{\rm days}$ beyond which all planets are dominated by insolation, with the exception of TOI-2096~c marked by the red circle. \textit{Middle:} The amplification in tidal heating in the presence of fluid tides as a function of the fluid dissipation frequency, $\sigma_{\rm R}$, is shown for each planet in the population (light curves), and averaged in the thick curves. The red curves assume spin-orbit synchronization, that is, the presence of eccentricity tides only, while the blue curves relax this assumption and allow for other tidal components. \textit{Right:} The resulting surface temperature,  assuming black body radiation, for a subset of 18 planets that are tidally dominated. Planets are assumed to be tidally locked into spin-orbit synchronization. Uncertainty bars in insolation-driven temperatures reflect the possible variations in planetary Albedo, while uncertainty bars in calculated temperatures allowing for tidal heating reflect variations in the magma ocean thickness and dissipation frequency.   }
\label{Fig_exo_heating}
\end{figure*}

\subsubsection{The effect of a magma ocean on the planetary energy budget}

In the first panel of Figure \ref{Fig_exo_heating}, we plot the ratio of tidal heating over insolation, $\mathcal{P}_{\rm T}/\mathcal{P}_{\rm ins}$,  for each planet, with and without a magma ocean. In both cases, the population follows a general trend which transitions from tidal dominance for ultra-short period planets to insolation dominance as the orbital period increases. As expected, this by virtue of insolation being $\propto a^{-2}$, while tidal heating being $\propto a^{-6}$ and decaying much faster. As such (with the single exception of a magma ocean coated TOI-2096 c, which is marked by the red circle), all planets in our population  beyond $P_{\rm orb}=4$ days are dominated by insolation. For $P_{\rm orb}\leq~4$ days, however, the number of planets whose thermal state is dominated by tides is dependent on whether they host a magma ocean or not, as we indicate with the probability distribution curves. Namely,  9 planets out of the 59 fall in the tidally dominated regime in the absence of fluid tides within a possible magma ocean, but the number increases to 18 if fluid tides are present.

The latter effect is due to the enhancement in heating by the fluid behavior in each planet, which we capture in the second panel of Figure \ref{Fig_exo_heating}. Therein, we compute for each planet in the list the tidal heating amplification if one accounts for fluid lava tides. We do so in two configurations: assuming spin-orbit synchronization, which in the absence of planetary obliquity leaves the planet subject to eccentricity tides only; and assuming asynchronous rotation, which amounts to additional tidal harmonics and therefore stronger tidal heating enhancement. While these close-in planets are expected to be in the 1:1 spin-orbit equilibrium, it is possible that they are  captured in higher order asynchronous states in the presence of thermal tides in their possibly existing atmospheres 
\citep[e.g.,][]{gold1969atmospheric,correia2001four,leconte2015asynchronous,auclair2017rotation,farhat2024thermal,revol2023spin,Valente24,Laskar_2024}{}{}. For both configurations, we also compute the average enhancement in tidal heating among the studied population. On average, fluid tides result in 1-2 orders of magnitude stronger tidal heating than pure solid tides, depending on the dissipation timescale in the magma ocean. Heating enhancement is stronger if the planet has nonzero eccentricity and is captured into an asynchronous equilibrium, reaching three orders of magnitude for certain planets. We emphasize here that we are not accounting for possible variations in tidal dissipation of the solid  mantle as we are interested in isolating the effect of tidal heating within a magma ocean. The effect of the former, pertaining to mantle temperature, viscosity, and the adopted rheology amounts to additional significant uncertainties in tidal heating \citep[e.g.,][]{renaud2018increased}{}{}.  

\subsubsection{Planetary surface temperatures}
Assuming blackbody radiation from the planet, we can then determine the equilibrium surface temperature, $T_{\rm s}$, controlled by the energy budget of each planet. Namely:
\begin{equation}
    T_{\rm s} = \left(\frac{\mathcal{P}_{\rm T}+P_{\rm ins}}{4\pi R_{\rm p}^2\epsilon\sigma_{\rm SB}} \right)^{1/4},
\end{equation}
where $\epsilon$ is the planetary emissivity \citep[0.9 in the infrared; e.g., ][]{henderson1996new}{}{}.  In the right panel of Figure \ref{Fig_exo_heating}, we show the resultant equilibrium surface temperature of the planets that fall in the tidally dominated region. We assume spin-orbit synchronization for each planet, but only from the perspective of tidal dynamics, that is, we do not model the possible significant  contrast in temperature between the dayside and the nightside of each planet. A caveat of our analysis in this setting is thus the assumption of a global magma ocean configuration, in contrast to the possibly more realistic hemispherical configuration present on the dayside of such planets \citep{boukare2022,meier2023}. However, the transition from the global to the hemispherical ocean configuration in the high friction regime brings about a slight offset in the tidal Love number, and consequently tidal heating \citep[e.g.,][]{farhat2022resonant,auclair2023can}{}{}, which can be effectively absorbed into the uncertainty of the ocean parameters $H$ and $\sigma_{\rm R}$. We compute the surface temperature in the case of: \textit{i)} insolation acting alone on the planet, \textit{ii)} insolation acting along with solid tides, and \textit{iii)} insolation added to solid mantle tides and fluid tides within a magma ocean. The uncertainty in the planetary albedo brings about the uncertainty bars in insolation-driven temperatures, while the uncertainties in tidally driven temperatures project variations of the magma ocean parameters. We emphasize that we have ignored additional thermal blanketing effects from possibly existing atmospheres. Our results can therefore be understood as demonstrating the effects of fluid tides in a conservative limit.

For a subset of the shown planets (55 Cancri\,e,  WASP-47\,e, Kepler-10\,b, and CoRoT-7\,b)  vigorous heating from the host star is most likely sufficient to partially melt the planets as the insolation-driven surface temperatures lie well above the typical melting points of silicates and even iron. For another subset (GJ 367\,b, TOI-1238\,b, and K2-138\,b), whether an insolation driven magma ocean is possible depends on the exact mantle phase diagram of each planet. However, with the additional contribution of solid tidal heating within the mantle, the three latter planets, along with TOI-1634\,b and L\,168-9\,b, become much more likely to harbour a magma ocean. The additional contribution of tidal heating is extreme in the case of TOI-1238\,b and TOI-1634\,b such that the resultant temperature is well above the typical liquidus of silicates and thus the existence of a magma ocean is guaranteed. Once a magma ocean is established on these planets, fluid tidal heating kicks in with an additional effect, driving the surface temperature to even higher values. The temperature then becomes significantly different from the equilibrium temperature usually calculated under the blackbody radiation assumption allowing for irradiation only (e.g., a difference of ${\sim}1400~{\rm K}$ for TOI-1238~b, ${\sim}1600{\rm K}$ for TOI-1634~b). For another subset of the planets (LHS~1140~c, LTT~1445A~c, L~98-59~b, GJ~486~b), the effect of tides is only considerable when one accounts for fluid tidal heating, and insolation-controlled temperatures are below silicate melting points. Namely, if magma oceans on these planets were ever formed, they can only persist by virtue of their intrinsic fluid tidal heating. Consequently, evidence of magma oceans on these planets, if any, would provide evidence on the significance of their fluid tidal heating. 

\subsubsection{Recent thermal emission measurements}

For the two planets with the shortest orbital period in our compiled population, GJ~367~b and 55~Cnc~e, a phase curve and secondary eclipse measurement, respectively, have been recently used to infer a brightness temperature. JWST phase curves of GJ 367 b indicate a higher dayside temperature ($\sim 1730$ K) than the maximum possible insolation-driven surface temperature \citep[assuming zero albedo and no heat redistribution;][]{zhang2024gj}{}{}. In the absence of a detectable atmosphere \citep[][]{zhang2024gj}{}{}, the observed brightness temperature can be considered to mirror the surface temperature of the planet, making this temperature discrepancy the first evidence, obtained from thermal emission spectra, on the effect of tidal heating on a sub-Earth. It is difficult, however, to indicate in this case whether this is the contribution of solid or fluid tidal heating due to the slight difference between their effects (Figure \ref{Fig_exo_heating}). 

In contrast, for 55~Cnc~e, although the fluid-driven temperature is well isolated and  significantly larger than the irradiation equilibrium temperature ($2131 \pm 156$ K), the possible presence of an atmosphere renders comparing the surface temperature of the planet to the observed brightness temperature inaccurate. Thermal emission spectra of the planet obtained between 2012 and 2013 using Spitzer indicate a dayside brightness temperature variation 
between $1365^{+219}_{-257}$ K and $2528^{+224}_{-229}$ K \citep[][]{demory2016variability,demory2016map}{}{}. More recent emission spectra captured by JWST indicate a brightness temperature of $1796\pm88$\,K \citep[][]{hu2024secondary}{}{}. This said, the variability in the planets brightness temperature extends to values that are: on one end, very high to be maintained by irradiation and solid tides only, potentially suggesting a significant contribution from fluid tidal heating in the magma ocean; and much lower than irradiation-driven temperatures on another end, suggesting the presence of a substantial secondary atmosphere which drives efficient heat redistribution. More interestingly, we anticipate that the observed spatial thermal variability in the dayside emission of the planet, and especially its asymmetry [\citet{demory2016map} reveals that the hotspot of the dayside is located $41\pm 12$ degrees east of the substellar point; though see \citet{mercier2022} for a contrasting analysis], can be explained by the effect of migrating fluid lava tides on the surface. Previous studies of the magma ocean dynamics on 55~Cnc~e \citep{meier2023} have been limited in their exploration of internal heating, thus a dedicated study of this planet's magma tides is certainly worthy of future efforts. The temporal variability of the planet's emission \citep{2023A&A...677A.112M} has been hypothesized to result from intermittent degassing from the interior \citep{2023ApJ...956L..20H}, which can explain the observed eclipse spectrum if the degassing is buffered by a large volatile reservoir in the magma ocean \citep{2023ApJ...954...29P}. Our present results add an important aspect to this debate: the effect of fluid tides in 55~Cnc~e raises the surface temperature to nearly $\sim$2800 K, suggesting an intermittent increase in refractory species in the atmosphere, which absorb in the shortwave. Thus, a fluid-tides-driven 55~Cnc~e would show variable abundances of UV absorbers in an otherwise volatile-dominated envelope. Further detailed magma ocean-climate studies are required to explore this hypothesis.

\section{Discussion and Outlook}\label{Section_discussion}
We have developed an analytical model describing the tidal response of a rocky planet harbouring a magma ocean. The modeled magma layer can be either residing on the surface of the planet, or underneath a thin solid  crust. We were drawn to explore this problem by the limitation of the tidal theories adopted in earlier works, where the magma medium is often  modeled as a low-viscosity solid. The aim of this work, therefore, is allowing for the fluid behavior of the magma ocean whenever its melt fraction exceeds a critical value, forcing a rheological transition. This is achieved by coupling the fluid dynamical equations in the {highly frictional} limit with the viscoelastic deformation equations describing the solid part of the planet. The earlier study of \cite{tyler2015tidal} was, to our knowledge, the only work in this direction, though it was focused on the fluid behavior of Io's putative magma ocean. Here, we expanded along these lines towards a more general planetary framework. The takeaway equation from our analytical formalism lies in \eq{eq_full_LN}, where we provide a closed-form solution for the Love number capturing the total tidal response of the planet. This frequency-dependent Love number carries the contribution of tides to the thermal state of the planet, and to the spin-orbit tidal evolution. 

We have shown that the fluid description of magma tides significantly changes the tidal response of the planet. Specifically, it amplifies the tidal response for a range of forcing frequencies, inducing a peak that is comparable in amplitude to that of the anelastic solid response, and that is otherwise not captured by the traditional treatment of magma tides (Figure \ref{Fig_Love_numbers}). Aiming at exploring this effect, we applied our developed formalism in two settings: the tidal evolution of the early Earth-Moon system (Section\,\ref{Section_Early_Earth_Moon}) and of close-in rocky exoplanets (Section \ref{Section_Close_In_Exo}). Before concluding on our findings, we draw first some remarks on the formalism and the examined applications. We start by discussing some limitations of our analysis, offering workarounds when possible and reporting on implications.

\begin{itemize}
\item[\textit{i)}] \textit{The case of a thick magma ocean:} In modeling fluid tides within the magma layer, we have utilized the Laplace Tidal Equations (LTEs) as the approximate set of equations describing momentum and mass tidal flow within an ocean of uniform depth residing on a rotating spherical planet. These equations, however, are valid under the assumption that the ocean is perfectly homogeneous, that is unstratified, and is relatively shallow such that the fluid's radial acceleration, along with the Coriolis acceleration corresponding to the horizontal component of the planet's rotation can be reasonably neglected. The major limitation in our developed formalism is thus when the magma ocean is relatively thick, as in the studied case of the early Earth when the mantle is, for some temperatures, fully molten (Section \ref{Section_Early_Earth_Moon}).

Evidently, one need not worry about the contribution of the Coriolis horizontal component in our setting as we are completely ignoring Coriolis effects in the creep flow formalism, in favor of the dominant dissipative contribution (see Section \ref{Section_Magma_isolated}). Nonetheless, in the case of a deep ocean, it has been realized that the LTEs would still hold if the ocean is stratified in density, whereby the uniform depth of the ocean, $H$, is replaced in the LTEs by a series of smaller equivalent depths \citep[see e.g.,][]{miles1974laplace,tyler2011tidal}{}{}. The solution of the horizontally varying quantities is then identical to that of the thin homogeneous layer case except that now the equivalent depths are obtained by solving an eigenvalue problem for the vertical structure. The latter then yields a dominant barotropic mode, which we capture in our formalism, along with a series of baroclinic modes corresponding to the series of equivalent depths.

This said, the worry in the adopted formalism lies in ignoring the vertically propagating internal wave modes that may  contribute to the surface tidal displacement, possibly altering our barotropic solution. However, these internal modes are simply inefficient in energy transport, and unless they are resonantly excited, they do not dominate over barotropic energetics \citep[e.g.,][]{wunsch1975internal,hendershott1981long,balmforth20052004}{}{}. To this end, in our highly dissipative magma medium, and in the absence of Coriolis, these modes can never be resonant. Besides, the contribution of these modes to the surface displacement is relatively very small, albeit controlling internal tidal flows. Therefore it is reasonable to claim that using the LTEs as governors of the tidal surface displacement in such settings is justified, and the obtained tidal surface displacement, which we capture in \eq{aux2}, serves as a first order solution to the problem, yet to be augmented with second order corrections from the baroclinic component.

\item[\textit{ii)}] \textit{The magma mush intermediate regime:} For the purpose of highlighting the effect of fluid tides within a magma ocean, we have modeled the temperature-driven rheological shift between the solid and liquid phases as an abrupt and discontinuous transition, both from a thermodynamic and a tidal perspective. Realistically, however, the transition occurs continuously and gradually, and it involves intermediate regimes of liquid flowing through a matrix of solid and crystal mush where the two phases coexist. Such biphasic regimes are suggested to describe the present state of Io's partially molten asthenosphere \citep[e.g.,][]{tyler2015tidal,bierson2016test}{}{}, a possible molten layer at the base of the Lunar mantle \citep[e.g.,][]{harada2014strong,walterova2023there}{}{}, the Martian mantle \citep[e.g.,][]{samuel2023geophysical}{}, and the low-density core of Enceladus \citep[e.g.,][]{roberts2015fluffy,choblet2017powering}{}{}. 

The tidal response of such mixed regimes cannot be described by the  viscoelastic gravitational theory of solid materials or the traditional treatment of fluid tides, thus a novel theory is required. The recent works of \citet{liao2020heat}, \citet{rovira2022tides}, and \citet{kamata2023poroviscoelastic}, based on the theory of poroviscoelasticity of \citet{biot1941general}{}{},  have made significant progress in this direction, whereby the classic deformation system of viscoelasticity is modified by coupling separate constitutive and dynamical equations for porous solids and pore liquids with Poisson's equation. The three works predict that a permeable flow of a pore fluid in a solid matrix, or a Darcian porous flow, produces a larger amount of heat than that yielded by solid deformation only. While it is expected that tidal heating would be increased by Darcy dissipation as the solid viscosity decreases with temperature and the pressure increases within the pores, we speculate that this enhancement in tidal heating smoothly follows the gradual transition between the rheological regimes, ultimately reaching maximal enhancement for a fluid tidal response, which we capture in this work. We thus anticipate the development of a complete model that self-consistently tracks the thermodynamic evolution along different turbulent-convective regimes, and the transition from a viscoelastic to poroviscoelastic rheological behavior, and eventually to a fluid  behavior and tidal response.     
\item[\textit{iii)}] \textit{On the temperature parameterization of the early Earth:} We used the temperature parameterization from \citet[][see \eq{convection_Soft_turbulence} and the discussion that followed]{korenaga2023rapid},  which is based on the assumption of a CO$_2$-H$_2$O-dominated atmospheric composition, derived using a grey atmospheric model. Consequently, variations in opacity as a function of atmospheric pressure and temperature were not accounted for. Similar to previous works in this area \citep[e.g.,][]{bower2022}, this modeling choice introduces a constraint on magma and outgassing composition limited to oxidized cases. Moreover, this model cannot reproduce climatic transitions, such as the runaway greenhouse effect. The focus on oxidized climates has been repeatedly corroborated by, for example, atmospheric \citep{pahlevan2019} and geochemical \citep{hirschmann2022} studies. The latter feature, however, may introduce unaccounted feedback effects in our model if the composition of the atmosphere undergoes significant changes during crystallization \citep[e.g., due to inhomogeneous crystallization or varied redox evolution,][]{2021ApJ...914L...4L}, which strongly affects the thermal blanketing effect \citep{lichtenberg2021}. This warrants follow-up studies with a radiative-convective climate model that accounts for the dynamic feedback effects between tidal and atmospheric forcing for the early Earth and in the context of short-period exoplanets.

\item[\textit{iv)}] {\textit{On the dissipative frequency} $\sigma_{\rm R}$: As indicated in Section \ref{Section_Magma_isolated}  and used in our model, we have characterized the dissipative damping timescale of the tidal flow by the frequency $\sigma_{\rm R}$ appearing as the drag term in the momentum equation. Nonetheless, whether energy in reality is dissipated through boundary friction, interior viscous resistance against flexure, form drag, or Darcy-like dissipation, $\sigma_{\rm R}$ characterizes the time required for the fluid to come to rest, under any of these dissipative mechanisms, after shutting off the tidal forcing. In this sense, one can relate $\sigma_{\rm R}$ to the more conventionally used tidal quality factor $Q$, when the latter is defined as a dimensionless timescale characterizing the attenuation of energy. Therefore, strictly under the definition of $Q$ as $2\pi$ times the ratio of energy stored to energy dissipated per tidal cycle, one can write: $Q=\sigma/(2\sigma_{\rm R})$. Our emphasis on the definition here arises from the fact that the usage of $Q$ is often misconstrued in the literature due to the varying definitions with implicit assumptions \citep[see for e.g., the conventional definition for fluid tides by][]{egbert2003semi}. In modeling fluid magma, and in the absence of constraints on this dissipative timescale, we have used $\sigma_{\rm R}$  as a free parameter, and we reported on the sensitivity of our results to its variations. Building on this work, however, independent constraints on $\sigma_{\rm R}$, along with physical parametrizations of the various dissipative mechanisms, are necessary in the future for a better understanding of the physics of lava tides. }
\end{itemize}

With these limitations in mind, we probed the effect of the fluid tidal response within magma oceans and learned the following:

\begin{itemize}
    \item In the case of the early Earth-Moon system, and in contrast with earlier studies, we show that magma tides drive efficient Lunar recession. Namely, the magma ocean is short-lived (a lifetime of $	\lesssim\,100$ kyrs), but enhanced fluid magma tides can expand the Lunar orbit to $\sim25 R_{\rm E}$ during its lifetime (Figure \ref{Fig_Earth_Moon_Distance}). It is important to emphasize that we have isolated here the effect of tides on a molten Earth in a geometrically coplanar-circular setting. The early Earth-Moon system, however, has much more to offer from a dynamical perspective. The system most probably underwent episodes of high eccentricity due to the evection resonance \citep[][]{touma1998resonances,ward2020analytical}{}{}, and possibly high mutual inclination in the case of a high obliquity Earth \citep[][]{cuk2016tidal,cuk2021tidal}{}{}. The system could thus be subject to significant eccentricity and obliquity tides that we have not examined in this work, and it would be interesting to see their effect in the fluid limit \citep[e.g.,][]{tyler2011tidal,chen2016tidal}{}{}.

    More importantly, we have not modeled Lunar tides, which in the presence of an analogous Lunar magma ocean, given our predictions here, should be more significant than predicted in earlier studies. Early Lunar tides act to oppose terrestrial tides \citep[e.g.,][]{touma1994evolution,daher2021long}{}{}, thus combining both terrestrial and Lunar tidal effects is necessary for constraining the rate of Lunar recession. This should then allow for  quantifying the probability of capture into the evection resonance \citep[][]{touma1998resonances}{}{}. The question of the lifetime of the Lunar magma ocean is also important for testing Lunar formation scenarios. Namely, the origin of the present $\sim5^\circ$ Lunar inclination to the ecliptic, and its reconciliation with the aftermath of the impact, is still debated \citep[e.g.,][]{touma1998resonances,ward2000origin,pahlevan2015collisionless,cuk2016tidal,tian2020vertical,downey2023thermal}{}{}; and the timing of the Lunar magma ocean solidification strongly affects the Lunar inclination evolution \citep[e.g.,][]{chen2016tidal}{}{}. To this end, our fluid tidal model for magma oceans can be efficiently used for both the Earth and the Moon, setting the table for  studying the rich interplay, in three dimensions, between these tidal players. 
    
    Moreover, we anticipate with our model to revive the study of the long term evolution of Venusian rotational dynamics, starting with the planet's formation. Namely, the seminal works of \citet[][]{correia2001four,correia2003long,correia2003long2}{}{} explored the possible rotational evolution histories of Venus, identifying the scenarios in which the current state of slow retrograde rotation can be explained. However, these studies were mostly considering initial rotational periods that are $\geq3~{\rm days}$. The latter constraint was imposed by the ability of Venusian tidal dissipation to decelerate the planet from its initial rotational period to the present state within the age of the Solar system. This said, the presence of an early Venusian magma ocean \citep[see e.g.,][]{widemann2023venus,salvador2023magma}{}{}, given the results we present here, can enhance the planet's tidal dissipation, allowing for the possibility of a more rapidly rotating young Venus. 

    \item In the case of close-in exoplanets, enhanced tidal heating from fluid magma flows affects their thermodynamic and spin-orbit states. We first showed how the  tidal spin-orbit evolution of a planet around its host star is affected: the typical chain of spin-orbit resonances that the planet undergoes on its way to spin-orbit synchronization is broken. This is by virtue of magma tidal heating dissipating orbital energy fast enough such that adiabatic capture into these resonance states is no longer feasible.  Consequently, the planet reaches tidal locking and orbital circularization over much shorter timescales. Interestingly then, planets with magma oceans have better chances of survival in multiplanetary systems than those without, as they are less prone to destructive gravitational interactions with other planets in the system, since they spend less time with eccentric configurations. 
    Finally, we have applied our tidal model to a population of detected close-in exoplanets whose surfaces are likely to be  molten. We showed how fluid tides render many of these exoplanets' surfaces dominated by tidal heating rather than vigorous insolation, and how this regime shift changes predictions for their surface equilibrium temperatures. 

    {Taking our model one step further, it would be interesting to study how fluid-induced tidal heating modifies the thermodynamic state of the  interiors of close-in planets, and specifically the equilibrium states that naturally arise in the competition between convective cooling and tidal heating \citep[e.g.,][]{moore2003tidal,moore2006thermal}. Namely, we anticipate that the effect of enhanced fluid tidal heating should allow the planet to sustain higher interior temperature equilibria,  which allow in turn for long-lived stable magma oceans, avoiding rapid and complete solidification or runaway melting of the whole planet \citep[e.g.,][]{peale1979melting,seligman2024potential}{}{}.  Interestingly, if it turns out that heating generated by fluid tides would allow the planet's mantle to occupy such very high temperature steady states, the increased longevity of the molten layer should enhance the capacity of short-period super-Earths to retain atmospheric volatiles by dissolution into magma \citep{lichtenberg2024}. This emphasizes the importance of such analysis for studies on the retention of secondary atmospheres \citep[e.g.,][]{hu2024secondary}. }

\end{itemize}

\section*{Acknowledgments}
The authors are grateful to the anonymous referee for thorough and insightful comments which helped improve the manuscript considerably. MF acknowledges support by the Munich Institute for Astro-, Particle and BioPhysics (MIAPbP), which is funded by the Deutsche Forschungsgemeinschaft (DFG, German Research Foundation) under Germany´s Excellence Strategy - EXC-2094 - 390783311, and where part of this work was completed.  TL acknowledges support by the Branco Weiss Foundation, the Alfred P. Sloan Foundation (AEThER project, G202114194), and NASA’s Nexus for Exoplanet System Science research coordination network (Alien Earths project, 80NSSC21K0593). This work has been supported by the French Agence Nationale de la Recherche (AstroMeso ANR-19-CE31-0002-01) and by the European Research Council (ERC) under the European Union’s Horizon 2020 research and innovation program (Advanced Grant AstroGeo-885250). This work was granted access to the HPC resources of MesoPSL financed by the Region Île-de-France and the project Equip@Meso (reference ANR-10-EQPX-29-01) of the  programme Investissements d’Avenir supervised by the Agence Nationale pour la Recherche. 
\appendix 
\section{Associated Legendre functions and Spherical Harmonics}\label{App_Legendre_Functions}
The convention used in the main text for the associated Legendre functions follows Chapter 8 of \citet[][]{abramowitz1988handbook}{}{} where the unnormalized functions read as:
\begin{equation} \label{def_alp}
    \bar{P}_n^m(\mu)= \frac{(-1)^m}{2^n  n!}(1-\mu^2)^{m/2}\partial_\mu^{n+m}(\mu^2-1)^n,
\end{equation}
which are solutions to the Legendre equation,
\begin{equation}\label{legendreeq}
   \partial_\mu\left[(1-\mu^2) \partial_\mu \bar{P}_n^m\right]+\left[n(n+1)-\frac{m^2}{1-\mu^2}\right]\bar{P}_n^m=0.
\end{equation}
The normalized associated Legendre functions used in the main text, ${P}_n^m$ satisfy:
\begin{equation}
    \int_{-1}^{1}{P}_n^m{P}_k^m = \delta_{n,k}.
\end{equation}
This implies the following normalization relation:
\begin{equation}
    {P}_n^m(\mu) = \left[ \frac{(2n+1)(n-m)!}{2(n+m)!}\right]\bar{P}_n^m(\mu).
\end{equation}
The definition of the spherical harmonics, normalized to one, follows from the latter definition of the normalized associated Legendre functions as:
\begin{equation}
    Y_n^m(\theta,\lambda) = \frac{1}{\sqrt{2\pi}}P_n^m(\cos\theta) e^{im\lambda}.
\end{equation}
\begin{figure}[]
\includegraphics[width=.45\textwidth]{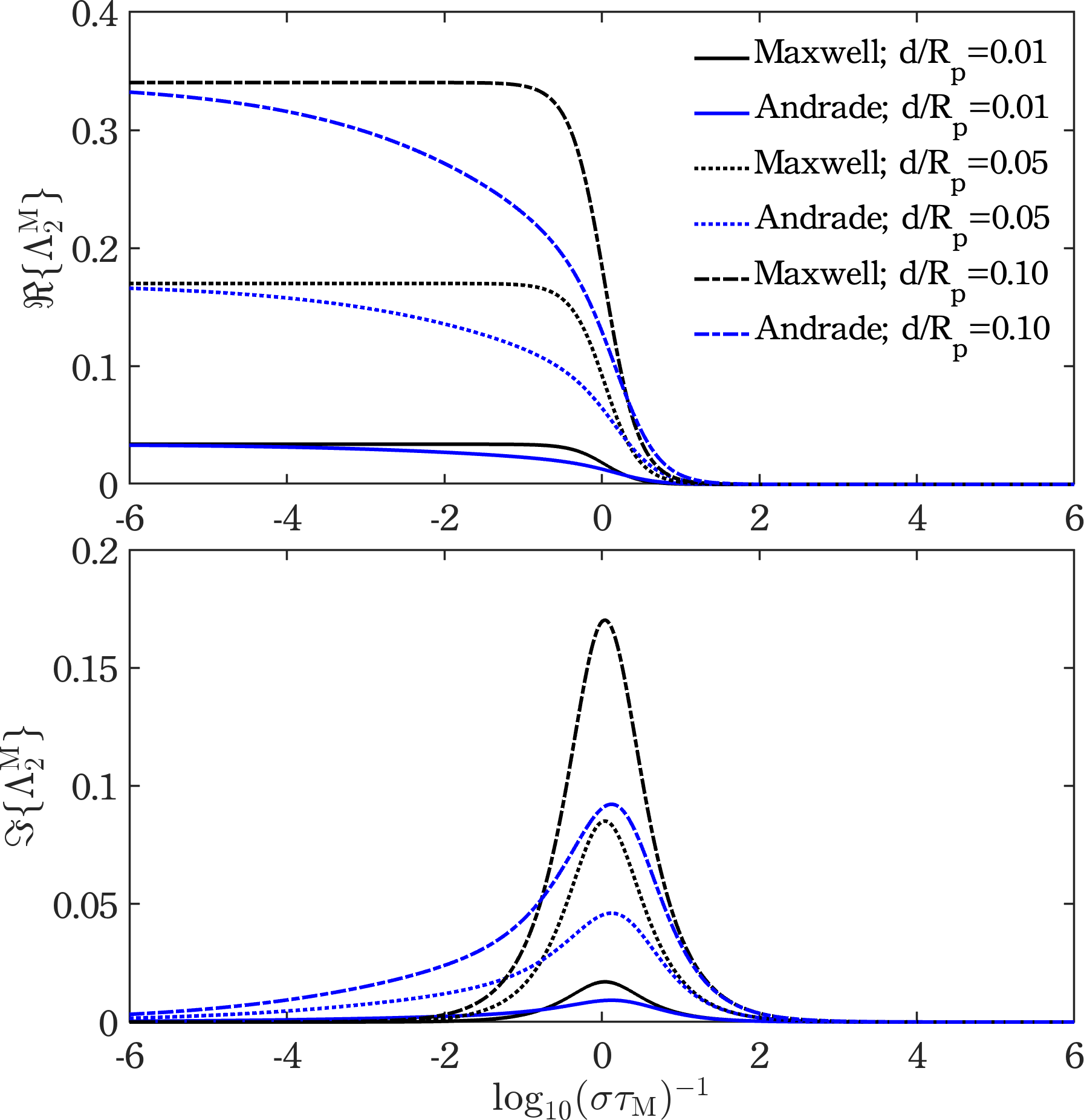}
\caption{The spectrum of the viscoelastic membrane constant of the crust. The real and the imaginary part of \eq{membrane_constant} are plotted  as a function of the inverse tidal forcing frequency or Maxwell timescale, for two different rheologies and three different thicknesses of the crust. Moving to the left of the x-axis is equivalent to increasing the viscosity of the crust, asymptotically reaching the limit of elastic isostatic compensation upon tidal forcing. In contrast, moving to the right places the crust in the limit of the response of an inviscid fluid.  }
\label{Fig_membrane_cst}
\end{figure}

\section{The viscoelastic membrane crust}\label{Appendix_Crust}
Here we provide the explicit expressions describing the viscoelastic response of the crust in the membrane limit assuming an Andrade rheology. The viscoelastic shear modulus for an Andrade rheology is given by \citep[e.g.,][]{efroimsky2012bodily}{}{}
\begin{equation}\label{shear_mod_andrade}
    \mu = \frac{\mu_{\rm E}}{1+(i\sigma \tau_{\rm A})^{-\alpha}\Gamma(1+\alpha) + (i\sigma\tau_{\rm M})^{-1}},
\end{equation}
where $\mu_{\rm E}$ denotes the elastic shear modulus. To obtain the Andrade Poisson ratio, $\nu$, we invoke the definition of the elastic bulk modulus $K_{\rm E} 	\equiv \frac{2}{3} \mu_{\rm E}(1+\nu_{\rm E})/(1-2\nu_{\rm E})$. In seismological models of the Earth's mantle anelasticity, bulk dissipation is found to be much smaller than shear dissipation \citep[e.g.,][]{widmer1991spherically,romanowicz20151}{}{}. For the Preliminary Reference Earth Model (PREM), for instance, the ratio for the mantle is $\leq0.1$ \citep[][]{dziewonski1981preliminary}{}{}. Therefore, for the purposes of modeling planetary dissipation, the bulk modulus is assumed to be real. This allows us to compute $\nu$ as a function of $\mu$, $\mu_{\rm E}$, and $\nu_{\rm E}$ and obtain
\begin{equation}
    \nu = \frac{\mu_{\rm E}(1+\nu_{\rm E})-\mu(1-2\nu_{\rm E})}{2\mu_{\rm E}(1+\nu_{\rm E}) + \mu(1-2\nu_{\rm E})}.
\end{equation}
\vspace{.3cm}

Using the latter equation with \eq{shear_mod_andrade}, we obtain the Poisson ratio for an Andrade rheology in the form:
\begin{equation}
    \nu  = \frac{3\nu_{\rm E} + (1+\nu_{\rm E})\left[ (i\sigma\tau_{\rm A})^{-\alpha}\Gamma ( 1+\alpha) + (i\sigma\tau_{\rm M})^{-1} \right]}{3 + 2(1+\nu_{\rm E})\left[ (i\sigma\tau_{\rm A})^{-\alpha}\Gamma ( 1+\alpha) + (i\sigma\tau_{\rm M})^{-1} \right]}.
\end{equation}
An analogous expression for a Maxwell rheology is obtained in \citet[][]{beuthe2015tides}{}{}. In Figure \ref{Fig_membrane_cst} we show the behavior of the membrane constant for Andrade and Maxwell rheologies, for different thicknesses of the crust, as a function of the crustal viscoelasticity, ranging from the elastic limit to the inviscid fluid limit.

\section{Close-in exoplanets population}

{
\startlongtable
\begin{deluxetable*}{lccccccc}
\tablenum{1}
\tablecaption{The studied population of detected Earth-like and Super-Earth exoplanets. }
\tablewidth{0pt}
\tablehead{
\colhead{Name} & \colhead{$M_{\rm p}(M_{\oplus})$} & \colhead{$R_{\rm p}(R_{\oplus})$} & \colhead{$P_{\rm orb}$ [days]} & \colhead{$e$} & \colhead{$M_\star(M_\odot)$} & \colhead{$R_\star(R_\odot)$} & \colhead{$T_\star[\rm K]$}}
\setlength{\tabcolsep}{5pt}  
\startdata
GJ 367  b & 0.546 & 0.577 & 0.322 & 0.06 & 0.45 & 0.46 & 3522 \\
55 Cnc e & 8.591 & 1.906 & 0.737 & 0.03 & 1.02 & 0.98 & 5196 \\
TOI-1238 b & 3.760 & 1.184 & 0.765 & 0.25 & 0.59 & 0.58 & 4089 \\
WASP-47 e & 6.830 & 1.767 & 0.790 & 0.03 & 1.04 & 1.14 & 5552 \\
Kepler-10 b & 4.608 & 1.449 & 0.837 & 0.06 & 0.91 & 1.06 & 5708 \\
CoRoT-7 b & 6.061 & 1.492 & 0.854 & 0.12 & 0.93 & 0.87 & 5313 \\
LHS 1678 b & 0.060 & 0.671 & 0.860 & 0.03 & 0.35 & 0.33 & 3490 \\
TOI-1634 b & 4.910 & 1.752 & 0.989 & 0.16 & 0.50 & 0.45 & 3550 \\
L 168-9 b & 4.608 & 1.361 & 1.402 & 0.21 & 0.62 & 0.60 & 3800 \\
GJ 486 b & 2.819 & 1.278 & 1.467 & 0.05 & 0.32 & 0.33 & 3340 \\
L 98-59 b & 0.413 & 0.832 & 2.253 & 0.10 & 0.32 & 0.29 & 3500 \\
K2-138 b & 3.115 & 1.536 & 2.353 & 0.40 & 0.93 & 0.86 & 5378 \\
K2-265 b & 6.541 & 1.679 & 2.369 & 0.08 & 0.92 & 0.98 & 5477 \\
HD 260655 b & 2.139 & 1.214 & 2.770 & 0.04 & 0.44 & 0.44 & 3803 \\
Kepler-21 b & 5.079 & 1.602 & 2.786 & 0.02 & 1.41 & 1.90 & 6305 \\
HD 38677 d & 3.350 & 1.364 & 2.882 & 0.07 & 1.21 & 1.26 & 6196 \\
TOI-554 c & 4.100 & 1.470 & 3.044 & 0.23 & 1.24 & 1.43 & 6338 \\
TOI-2096 b & 1.907 & 1.217 & 3.119 & 0.15 & 0.22 & 0.25 & 3285 \\
LTT 1445A c & 1.541 & 1.123 & 3.124 & 0.22 & 0.26 & 0.27 & 3340 \\
TOI-1695 b & 5.784 & 1.866 & 3.134 & 0.10 & 0.54 & 0.53 & 3630 \\
K2-199 b & 6.897 & 1.693 & 3.225 & 0.02 & 0.73 & 0.68 & 4648 \\
TOI-270 b & 1.481 & 1.221 & 3.360 & 0.02 & 0.40 & 0.38 & 3386 \\
K2-314 b & 8.750 & 1.909 & 3.595 & 0.06 & 1.05 & 1.71 & 5430 \\
L 98-59 c & 2.218 & 1.356 & 3.691 & 0.10 & 0.32 & 0.29 & 3500 \\
LHS 1678 c & 0.391 & 0.921 & 3.694 & 0.04 & 0.35 & 0.33 & 3490 \\
Kepler-20 b & 9.700 & 1.833 & 3.696 & 0.03 & 0.95 & 0.96 & 5495 \\
LHS 1140 c & 1.808 & 1.255 & 3.778 & 0.31 & 0.15 & 0.19 & 3131 \\
TOI-836 b & 4.529 & 1.668 & 3.817 & 0.05 & 0.68 & 0.67 & 4552 \\
GJ 357 b & 2.087 & 1.142 & 3.931 & 0.05 & 0.34 & 0.34 & 3505 \\
HD 23472 d & 0.550 & 0.734 & 3.977 & 0.07 & 0.75 & 0.73 & 4813 \\
K2-415 b & 2.860 & 0.994 & 4.018 & 0.03 & 0.16 & 0.20 & 3173 \\
HD 22946 b & 2.609 & 1.333 & 4.040 & 0.13 & 1.10 & 1.16 & 6210 \\
TOI-1136 b & 3.496 & 1.866 & 4.173 & 0.03 & 1.02 & 0.97 & 5770 \\
K2-32 e & 2.098 & 0.988 & 4.349 & 0.04 & 0.86 & 0.84 & 5315 \\
LHS 1678 d & 0.918 & 0.960 & 4.965 & 0.04 & 0.35 & 0.33 & 3490 \\
K2-111 b & 5.289 & 1.778 & 5.352 & 0.13 & 0.84 & 1.25 & 5775 \\
LTT 1445A b & 2.870 & 1.278 & 5.359 & 0.11 & 0.26 & 0.27 & 3340 \\
HD 38677 e & 4.128 & 1.440 & 5.516 & 0.07 & 1.21 & 1.26 & 6196 \\
HD 260655 c & 3.089 & 1.501 & 5.706 & 0.04 & 0.44 & 0.44 & 3803 \\
pi Men c & 3.630 & 1.835 & 6.268 & 0.15 & 1.09 & 1.10 & 6037 \\
TOI-2096 c & 4.608 & 1.874 & 6.388 & 0.10 & 0.22 & 0.25 & 3285 \\
HD 38677 c & 9.598 & 1.776 & 6.584 & 0.06 & 1.21 & 1.26 & 6196 \\
Kepler-100 b & 7.310 & 1.295 & 6.887 & 0.13 & 1.08 & 1.49 & 5825 \\
Kepler-1876 b & 2.415 & 0.835 & 6.992 & 0.01 & 1.19 & 1.48 & 6104 \\
L 98-59 d & 1.939 & 1.489 & 7.451 & 0.07 & 0.32 & 0.29 & 3500 \\
HIP 113103 b & 5.200 & 1.649 & 8.145 & 0.02 & 0.58 & 0.70 & 4800 \\
K2-266 c & 0.289 & 0.691 & 7.814 & 0.04 & 0.69 & 0.70 & 4285 \\
HD 23472 e & 0.721 & 0.801 & 7.908 & 0.07 & 0.75 & 0.73 & 4684 \\
K2-21 b & 1.980 & 1.800 & 9.003 & 0.10 & 0.68 & 0.65 & 4222 \\
Kepler-338 e & 8.581 & 1.525 & 9.341 & 0.05 & 1.10 & 1.74 & 5923 \\
TOI-700 b & 1.290 & 0.895 & 9.977 & 0.08 & 0.42 & 0.42 & 3480 \\
Kepler-11 b & 1.907 & 1.767 & 10.304 & 0.05 & 0.96 & 1.06 & 5663 \\
Kepler-197 c & 5.403 & 1.207 & 10.350 & 0.08 & 1.09 & 1.12 & 6004 \\
HD 136352 b & 4.678 & 1.609 & 11.578 & 0.08 & 0.87 & 1.06 & 5664 \\
HD 23472 f & 0.769 & 1.113 & 12.162 & 0.07 & 0.75 & 0.73 & 4684 \\
KOI-1599.02 & 9.001 & 1.866 & 13.609 & 0.01 & 1.02 & 0.97 & 5370 \\
Kepler-138 c & 1.971 & 1.174 & 13.781 & 0.05 & 0.52 & 0.44 & 3841 \\
HD 23472 b & 8.320 & 1.833 & 17.667 & 0.07 & 0.75 & 0.73 & 4684 \\
TOI-2095 b & 4.100 & 1.224 & 17.775 & 0.12 & 0.47 & 0.45 & 3662 \\
\enddata
\tablecomments{The population of 59 planets was compiled from \url{http://exoplanet.eu} in April 2024. We constrained the available catalogue at the time to planets with masses $M_{\rm p}\leq10 M_{\oplus},$ radii $ R_{\rm p}\leq 2R_{\oplus}$, orbital periods $P_{\rm orb}\leq 20$ days, and non-zero eccentricity. It should be emphasized that values of the latter, specifically, are nominal and subject to significant uncertainties. Besides the planetary parameters, the stellar hosts' masses $(M_{\star})$, radii $(R_{\star}$), and effective temperature ($T_{\star}$) are  shown. The planets are sorted by their orbital periods.  }
\label{Table_Exo}
\end{deluxetable*}}



\end{document}